\documentclass[a4paper,11pt]{article}
\pdfoutput=1 

\usepackage{jheppub} 

\usepackage[utf8]{inputenc}

\usepackage{tikz}

\usepackage{amsmath,amssymb,amsthm}

\usepackage{graphicx}
\usepackage{wrapfig}
\usepackage{framed}
\usepackage{todonotes}

\graphicspath{{Bilder/}}

\title{Heavy Quarks in Strongly Coupled Non-Conformal Plasmas with Anisotropy}

\author[a]{Enrico Brehm}

\affiliation[a]{Max-Planck-Institut für Gravitationsphysik, Albert-Einstein-Institut, \\Am Mühlenberg 1, D-14476 Golm, Germany}

\emailAdd{brehm@aei.mpg.de}

\abstract{We analysis physical observables of heavy quarks in gravity models describing strongly coupled non-conformal plasmas with anisotropy via the gauge/gravity duality. The focus lies on the binding energy of static quark-antiquark $(q\bar q$-)pairs, the maximum distance (screening distance) of a bound $q\bar q$-pair and the drag force on uniformly moving quarks in the hot plasma. In order to discover universal behavior in the observables, the computations are worked out in a two parameter deformation of pure gravity in $AdS_5$ spacetime with a black brane which is assumed to be dual to a respective two parameter deformation of $\mathcal{N}$=\,4 supersymmetric Yang-Mills (SYM) theory at temperature $T$. The deformation is designed to break isotropy and conformal symmetry and is a solution to equations of motion of a gravity action. }

\renewcommand{\t}{\text{t}}
\renewcommand{\c}{\text{c}}
\newcommand{\SYM}{$\mathcal{N}$=\,4 SYM theory}
\renewcommand{\sc}{\text{s}}

\begin{document} 
\maketitle
\flushbottom

\section{Introduction}

The discovery of the AdS/CFT correspondence in the late 1990s \cite{Maldacena:1997re}, and its generalizations now referred to as the gauge/string duality, have led to a broad research field in which one studies the relation of string theory to the strong-coupling limit of a large class of quantum field theories in lower dimensions. It can for example help us to a better understanding of the strong interaction, i.e. quantum chromodynamics (QCD).

The gauge/string duality may for example be used for the description of a hot plasma of particles in a gauge theory. A concrete example for such a state of matter is the Quark--Gluon Plasma (QGP) which is a phase of QCD that appears for temperatures higher than $T_\text{c} \approx 160$ MeV. There quarks and gluons can propagate independently and are no longer bound in colour-neutral hadrons \cite{Stephanov:2007fk}. On earth, the QGP can be produced in ultrarelativistic heavy ion collisions. It is widely believed that RHIC and LHC succeeded in producing the QGP and that for temperatures not too far above $T_\text{c}$ it behaves as a fluid, i.e. it is a strongly coupled system \cite{Arsene:2004fa,Back:2004je,Adams:2005dq,Adcox:2004mh,Aamodt:2010pa,ATLAS:2011ah,Chatrchyan:2012ta} (for theoretical reviews see \cite{Gyulassy:2004vg,Gyulassy:2004zy,Muller:2007rs}).

In general, there exist several theoretical methods that are used to get more insight into the rich structure of QCD at high temperature. There are for example the Dyson--Schwinger equations \cite{Dyson:1949ha,Schwinger_1951a}, chiral perturbation theory \cite{Gasser:1984gg}, heavy quark effective field theories \cite{Neubert:2005mu,Sommer:2010ic}, the functional renormalization group approach \cite{Wetterich:1992yh}, and lattice QCD \cite{PhysRevD.10.2445}. The gauge/string duality is an additional promising tool to  describe strongly coupled systems like the QCP. It includes systems at finite temperature by considering Schwarz\-schild-type black branes \cite{Aharony:1999ti,Witten:1998zw} and -- by the nature of the duality -- it maps the strong coupled regime of a gauge theory to the weakly coupled regime of the dual string theory, s.t. calculation can be performed in a \textit{classical} (super)gravity theory. 

To describe the physics of the QGP with the help of the gauge/string duality one needs to find the dual string theory description of QCD, which is so far not in sight. Nevertheless, one can see promising progress in emulating the thermodynamics of high-temperature QCD and related physics with the help of the gauge/string duality \cite{Arefeva:2016rob}.  

Roughly speaking there exist two philosophies within the community of gauge/string duality applied to hot plasma. The first philosophy follows the idea to explicitly construct the holographic dual description of QCD or something close to it. Within this approach one for example computes meson spectra or the running of the coupling and compares to known results. Either one uses a \textit{top-down} construction where one starts from ten-dimensional string theory and tries to \textit{derive} the requested dual theory (see for example  \cite{Polchinski:2000uf,Karch:2002sh,Sakai:2004cn}) or a \textit{bottom-up} construction where one usually starts with a five-dimensional supergravity action with a matter content of fields and solves their equations of motion (see for example \cite{Gursoy:2007cb,Galow:2009kw}).

The second philosophy follows the \textit{universality approach}. Here, one focuses on larger classes of theories and so tries to understand the generic physics of strongly coupled theories (see for example \cite{Natsuume:2007qq,Springer:2009tu,Ewerz:2010du}). The goal is to find universal behavior of certain physical phenomena or observables in these classes. The larger the class of theories the stronger is the significance of these universalities. The hope is that one can extend these generic results to QCD, even though QCD is still very different from all gauge theories that possess a known holographic dual description. Possibly the most famous result within this approach is the lower bound in a vast class of theories for the ratio of the shear viscosity $\eta$ of a strongly coupled plasma to the entropy density $s$ given by $\eta/s \ge 1/(4\pi)$, which is known as the \textit{KSS bound} \cite{Kovtun:2004de}.\footnote{This bound named after Kovtun, Son and Starinets is, however, not true for anisotropc holographic models as e.g. shown in \cite{Samanta:2016pic,Chakraborty:2017msh,Rebhan:2011vd}.} 

Here, we follow the second philosophy, namely to construct new classes of models and analyse several observables within these classes. 

As mentioned before, the QGP develops after the ultrarelativistic collision of heavy ions. The resulting plasma is strongly coupled and has finite temperature, both properties treatable within the gauge/string duality. However, there are more features which need to find their description within the duality to result in a proper dual theory for the QGP. First, the theory must not be conformal. In addition to that, the generic QGP produced in collisions is inhomogeneous, of finite size, and is manifestly anisotropic.\footnote{The latter is true for a short time after the collision and has been studied from a hydrodynamical perspective e.g. in \cite{Florkowski:2008ag,Florkowski:2009sw,Ryblewski:2010tn,Florkowski:2010cf,Martinez:2010sc,Ryblewski:2010bs,Ryblewski:2010ch,Biondini:2017qjh}. It is also worth mentioning that not only in the context of the QGP but also in the holographic description of superconductivity anisotropy is under investigation \cite{Bai:2014poa,Koga:2014hwa}.}

In this text, we focus on the non-conformal and anisotropic aspects. Starting from \SYM, which is dual to pure AdS space, we first show the construction of two single-parameter deformations that respectively break conformal invariance and isotropy. We then in particular combine the feature of anisotropy and non-conformality in one framework. This will help us to investigate a larger class of anisotropic theories and gives us more insight in the nature of anisotropic, non-conformal, strongly coupled plasmas. 

The first one-parameter deformation is a bottom-up constructions breaking conformal invariance even at zero temperature by coupling a scalar field to pure gravity in AdS. The second deformation is a top-down construction of an anisotropic deformation that was first introduced in \cite{Mateos:2011ix,Mateos:2011tv} (Other versions of anisotropic deformations of \SYM~ are e.g. \cite{DHoker:2009mmn,Rougemont:2015oea}, where anisotropy is due to strong magnetic fields generated in the overlapping region of the colliding nuclei, or \cite{Donos:2016zpf}). We will then combine the anisotropic with a non-conformal model at the supergravity level. This is not generically possible for any two one-parameter deformations. We solve the equations of motion for a good range of the two parameters. Hence, we construct a class of non-conformal, anisotropic models specified by these two parameters. 

After the construction we investigate various observables in the two-parameter model. We will have a short look at the thermodynamics, especially at the dependency of the entropy density and the temperature of the plasma on the parameters that specify the deformation. We then focus on observables of heavy quarks surrounded by the anisotropic plasma. We compute the screening distance which is the maximally possible distance between two quarks so that they can still form a bound state. For distances larger than that the quarks are said to be screened from each other by the plasma.  We also compute the binding energy between the two particles for all distances smaller than the screening distance. At last we compute the drag force experienced by a single quark moving with constant velocity through the plasma. We will see that all the observables will behave differently under (i) isotropic non-conformal and (ii) anisotropic deformations. We will also see that the stronger the anisotropic deformation the stronger do the different observables depend on orientations relative to the anisotropic direction. 

We organized the text as follows. In section \ref{sec:constructions} we show the construction of the three previously mentioned models. In the following section \ref{sec:observables} we concern ourself with the derivation of the observables. We show the general methods to derive the respective quantities and then apply them to our two-parameter deformation. We summarize the main results and give a general conclusion of this paper in chapter \ref{sec:summary}. Finally, some technical derivations are presented in the appendices. 

\section{\boldmath Deformations of $\mathcal N$=\,4 SYM theory}\label{sec:constructions}

\subsection{Single-Parameter Isotropic Deformation}
\label{sec:deform1}

In this section we introduce a one-parameter deformation of AdS$_5$ space. We assume that the deformation on the gravity side also leads to a deformation of the dual \SYM~and in particular breaks conformal invariance explicitly. It shows similar effects on observables as other known non-conformal deformations, e.g. in \cite{Ewerz:2010du}. It will turn out that the present model can consistently be combined with the upcoming anisotropic deformation which is the main reason we present it here. 

The deformation is a bottom-up construction solving 5D gravity equation of motions with specific boundary conditions. The starting point is a gravity action of the form
\begin{equation}
 S = \frac{1}{16\pi G_5}\int d^5x\,(\mathcal{L}_\text{gravity} + \mathcal{L}_\text{matter})\,, \label{eq:isoactio}
\end{equation}
with $\mathcal{L}_\text{gravity} = \sqrt{-g} (\mathcal{R}- 2\Lambda)$, where $\mathcal{R}$ is the Ricci scalar, $\Lambda \equiv-6$ is the cosmological constant, and $G_5$ is the five dimensional Newton constant. In our calculations we choose the AdS radius to be one, $L_{AdS} \equiv 1$. For a vanishing matter content, $\mathcal{L}_\text{matter} = 0$, solving the latter action gives pure Anti-de-Sitter (AdS) space, that is dual to $\mathcal{N} = 4$ SYM theory. Our deformation Ansatz is to introduce a single scalar $\tau(u)$ that depends solely on the additional holographic coordinate $u$ and, among other features, breaks conformal invariance on the level of the action. In addition to a kinetic term, the scalar comes with a potential that we choose to be
\begin{equation}
 V(\tau) = -2 \tau^2 -c \tau^4\,. 
\end{equation}

\noindent
It fixes the squared mass of the scalar at $m^2=-4$, which is exactly the Breitenlohner-Freedman bound for consistent scalar solutions in AdS$_5$-like gravitational backgrounds \cite{Breitenlohner:1982bm}. As we will see, for suitable boundary conditions the pre-factor $c$ parametrizes the deformation, where we regain the AdS-background for $c=0$. The choice for the potential is somewhat arbitrary. The guiding idea is to construct a simple  model that can be combined with the upcoming anisotropic deformation.  The gravity action is thus given by
\begin{equation}
 S = \frac{1}{16 \pi G} \int d^5x \sqrt{-g} \left(\mathcal{R} + 12 - \frac{1}{2}(\partial \tau)^2 + 2 \,\tau^2 + c \,\tau^4\right).
\end{equation}

\noindent
To solve this we have to specify a metric Ansatz. We assume a metric that is diagonal, invariant under translations in all boundary coordinates $(t,\vec{x})$ and under rotations in the spatial boundary coordinates $(\vec{x})$ for all $u$. We assume a temperature so that boost symmetry is broken. A most general metric Ansatz under these conditions is given by
\begin{figure}[t]\centering
 \includegraphics[width=0.49\textwidth]{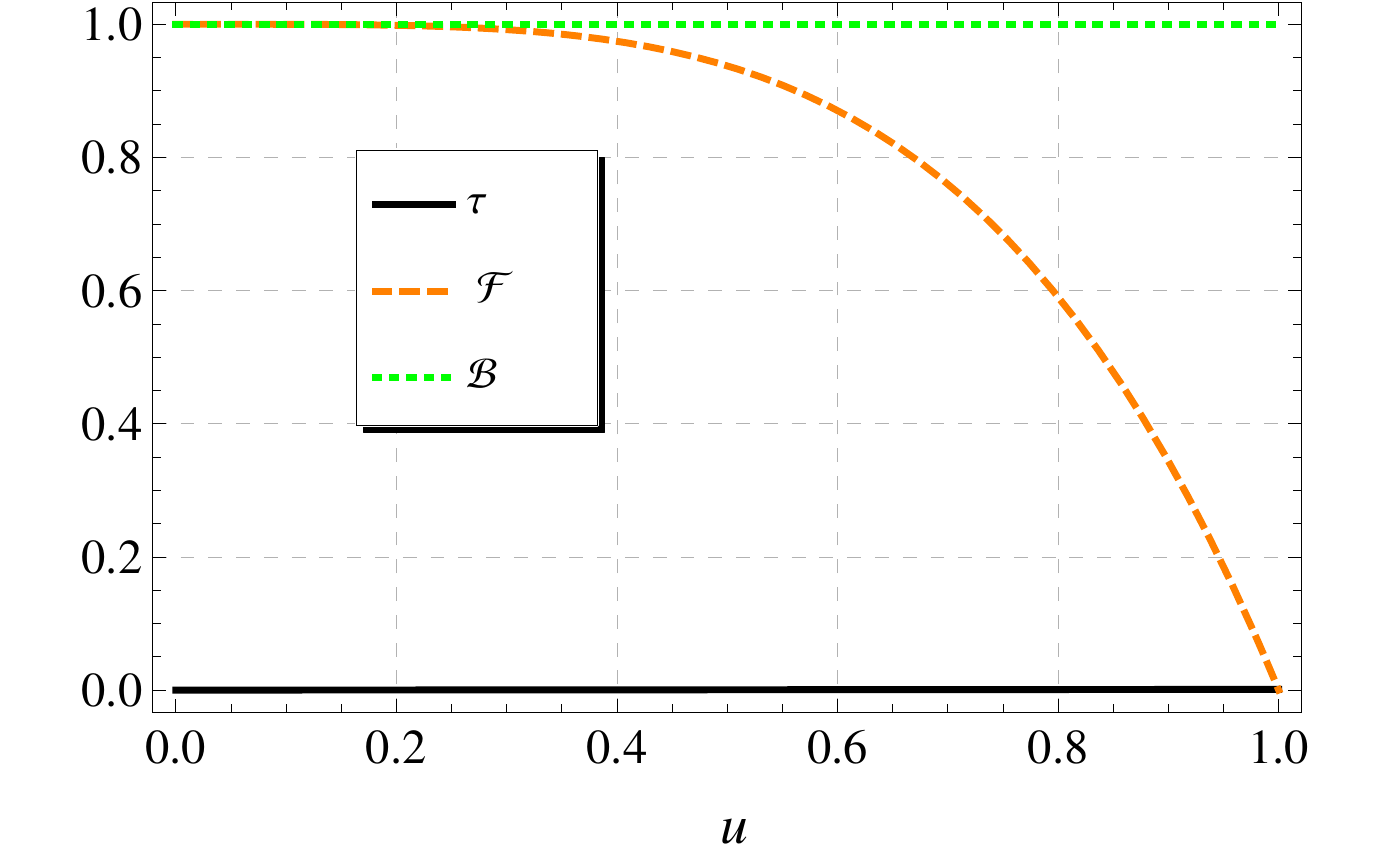}
 \includegraphics[width=0.49\textwidth]{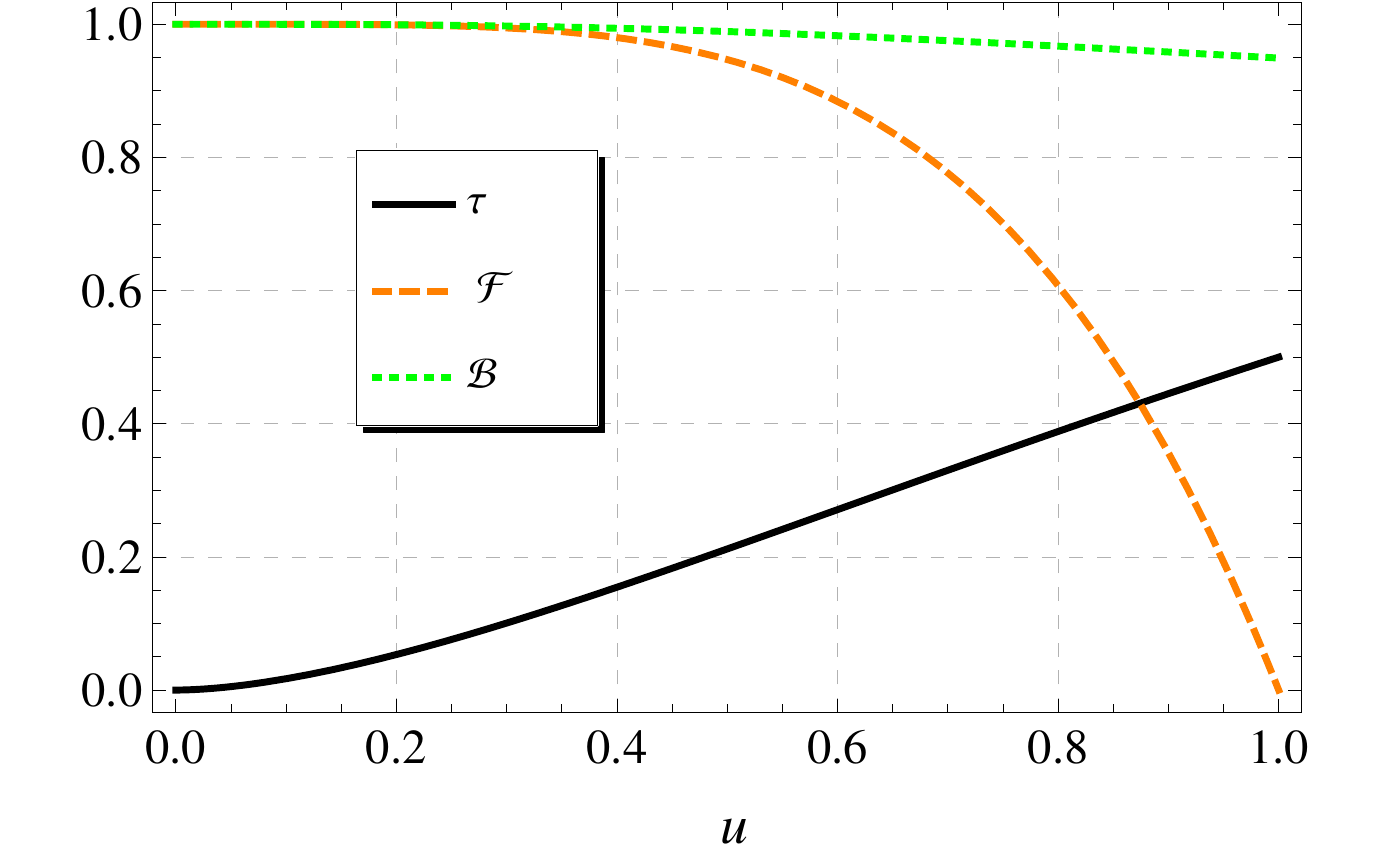}
 \includegraphics[width=0.49\textwidth]{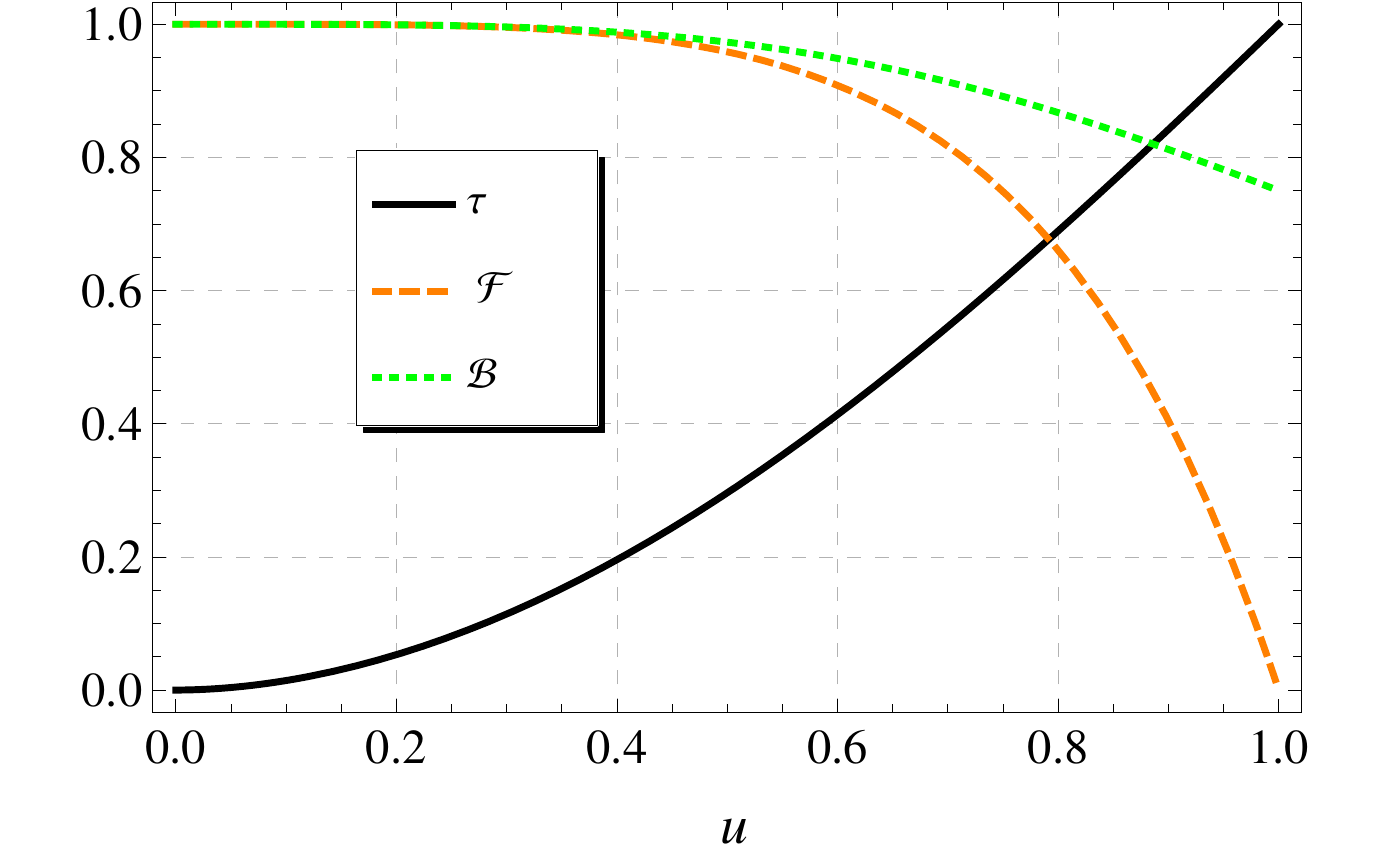}
 \includegraphics[width=0.49\textwidth]{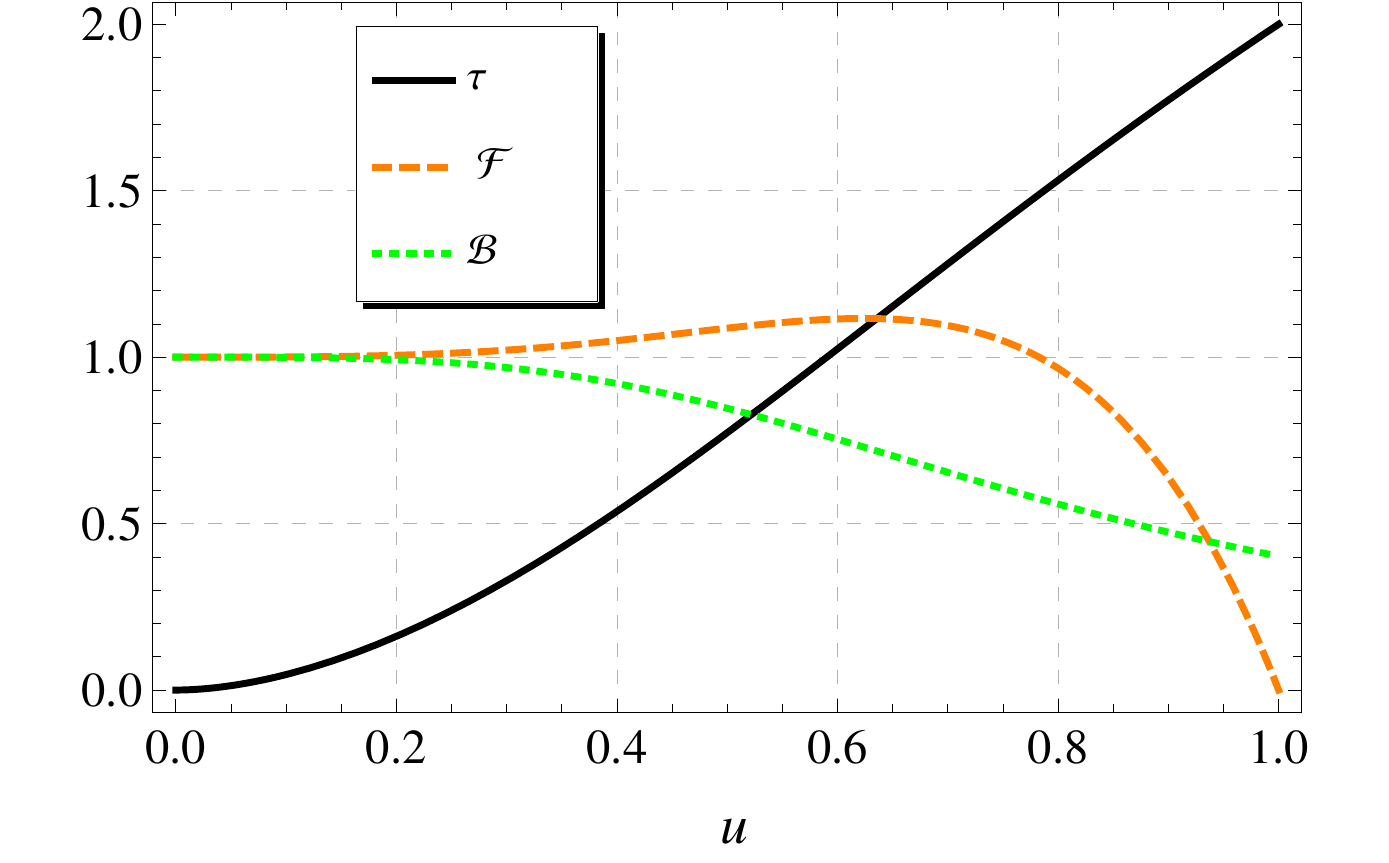} 
 \caption{The solutions for $\tau,~\mathcal{F}$ and $\mathcal{B}$ for $u_\text{h} = 1$ and $c=0$ (top left), $c=0.5$ (top right), $c=1$ (bottom left) and $c=2$ (bottom right).}
 \label{fig:solDeform2}
\end{figure}
\begin{equation}
 ds^2 = \frac{1}{u^2}\left(-\mathcal{FB}\, dt^2 +dx^2 + dy^2 + dz^2 +\frac{du^2}{\mathcal{F}}\right),\label{eq:IsoMetric}
\end{equation}
where $\mathcal{F,B}$ are functions of the holographic coordinate $u$ only. Some algebra shows that the equations of motions for the action \eqref{eq:isoactio} can be reduced to the two simple differential equations for the metric functions  
\begin{align}
 \label{eq:Bdeform2}\frac{\mathcal{B}'}{\mathcal{B}}=\,& -\frac{u \,(\tau')^2}{3}\,,\\
 \label{eq:Fdeform2} \mathcal{F} = \,&\frac{-12(\tau +c\tau^3)+u(12+2\tau^2+c\tau^4)\tau'}{3 u (\tau' +u \tau'')}
\end{align}
and a third order differential equation for $\tau$ given by \eqref{eq:tauDGL2} in appendix \ref{app:equ}. It can be solved numerically for given boundary conditions. At first we demand the existence of a horizon $u_\text{h}$, so $\mathcal{F}(u_\text{h}) = 0$. Next the scalar should vanish at the boundary, so $\tau(0) = 0$. The last boundary condition arises from the requirement to recover AdS for vanishing parameter $c$. One simple choice that fulfills this requirement is $\tau(u_\text{h}) = c$, s.t. the scalar vanishes at the horizon for $c=0$. We then choose the trivial solution for $\tau$, i.e. it vanishes everywhere in the bulk. In that case $\mathcal{L}_\text{matter} = 0$ and we recover the black brane AdS$_5$ solution. Other choices of boundary conditions for $\tau$ at $u_h$ are possible, as long as $\tau(u_h) = 0$ for $c=0$. However, we take the simple one from above.   

With the specification of the boundary conditions the construction above defines a one-parameter deformation of AdS$_5$. It can be solved numerically for a large interval of the deformation parameter $c$, so that it is well suited to investigate a large number of theories. In figure \ref{fig:solDeform2} we visualize the solutions for $\mathcal{F,B},\tau$ for several values of $c$. Already there we see how the difference of the functions to pure AdS space grows for larger $c$.

\subsection{Single-Parameter Anisotropic Deformation}
\label{sec:MT}
Before we come to the explicit construction let us shortly state some general considerations. We want to consider models with geometries that are dual to uniform, static and anisotropic plasmas. A corresponding stress-energy tensor of the plasma should look like 
\begin{equation}
 \label{eq:EnergyMomentumTensor}
 \langle T_{\mu\nu}\rangle = \begin{pmatrix}
  \varepsilon & 0 & 0 & 0\\
  0 & p_\perp & 0& 0 \\
  0 &0 &  p_\perp & 0 \\
  0  & 0 & 0 & p_\parallel
 \end{pmatrix}\,,
\end{equation}
where $p_\perp$, $p_\parallel$ are the pressures perpendicular and parallel to the anisotropic direction, and $\varepsilon = 2p_\perp + p_\parallel$ is the energy density of the plasma (see e.g. \cite{Janik:2008tc}). For a uniform and static plasma all the entries of the energy momentum tensor have to be constants in both time and space. We assume that the 5D dual gravitational theory has the same symmetries as the stress energy tensor. Then a general metric Ansatz for that theory is given by
\begin{equation}
 \label{eq:OurMetricAnsatz}
 ds^2 = \frac{1}{u^2}\left(-\mathcal{FB}~ dt^2 +dx^2 + dy^2 + \mathcal{H} dz^2 +\frac{du^2}{\mathcal{F}} \right)\,, 
\end{equation}
where $\mathcal{F,H}$, and $\mathcal{B}$ are functions of the holographic coordinate $u$ only. 

We now shortly want to present an anisotropic top down construction -- a solution to type IIB supergravity equations of motions. It was introduced in \cite{Mateos:2011tv,Mateos:2011ix} where they were motivated by string theory duals of Lifshitz-like fixed point \cite{Azeyanagi:2009pr}.  

The field theory side is a spatially deformed version of $\mathcal{N} = 4$ finite temperature SYM theory. The deformation term is included in the action and contains a $\theta$-parameter that depends linearly on one of the three spatial coordinates, $\theta = 2 \pi n_{D_7} z$, where $(t,x,y,z)$ are the coordinates of the gauge theory. $n_{D_7}$ is a constant that can be interpreted as a density of $D_7$-branes distributed along the $z$-direction. The total gauge theory action then takes the form  
\begin{equation}
 S_\text{gauge} = S_{\mathcal{N} = 4} + \delta S ~~~~\text{with}~~~~\delta S  = \frac{1}{8 \pi^2} \int \theta(z) \text{Tr}(F\wedge F).
\end{equation}

\noindent
In the gravity dual $\theta$ is related to the axion $\chi$ of type IIB supergravity through the complexified coupling
\begin{equation}
 \xi = \frac{\theta(z)}{2 \pi} + \frac{4 \pi i}{g_\text{YM}^2} = \chi(z) + ie^{-\phi}\,,
\end{equation}
where $\phi$ is the dilaton and $\chi$ the axion, which can be shown to be of the form $\chi = a z$. This is because the axion is magnetically sourced by the $D_7$-branes distributed along the $z$-direction, and in particular $a\propto n_{D_7}$. The larger $a$ the stronger the branes back-react on the geometry anisotropically. In this sense one can see $a$ as an anisotropy deformation parameter. 

The ten-dimensional solution to this construction is a direct product $\mathcal{M}\times S^5$, where $S^5$ is the five-sphere. Now $\mathcal{M}$ can be seen as a solution of a 5D supergravity action whose excited fields are the 5D metric $g_{\mu\nu}$, the axion $\chi$ and the dilaton $\phi$.  Their dynamics is governed by the five-dimensional axion-dilaton-gravity action which in Einstein frame takes the form \cite{Mateos:2011tv}
\begin{equation}
 \label{eq:AxionDilatonAction}
 S = \frac{1}{16\pi G_5} \int_\mathcal{M} d^5x \sqrt{-g} \left(\mathcal{R} +12 -\frac{1}{2} (\partial \phi)^2 -\frac{1}{2}e^{2\phi}(\partial\chi)^2\right) + \text{boundary terms}\,,
\end{equation}
where the dilaton $\phi$ only depends on $u$. Note that in Einstein frame the metric Ansatz \eqref{eq:OurMetricAnsatz} comes with an additional overall factor of $e^{-\frac{\phi}{2}}$.

Deriving the equations of motion of \eqref{eq:AxionDilatonAction} one finds that the functions $\mathcal{F,B}$ and $\mathcal{H}$ can be given in terms of the dilaton as
\begin{align}
\label{eq:H} \mathcal{H} &= e^{-\phi}\,,\\
 \mathcal{F} &=\frac{e^{-\frac{\phi}{2}} }{4 \left(u \phi
   ''+\phi'\right)}\left(a^2 u e^{\frac{7 \phi}{2}}
   \left(u \phi'+4\right)+16 \phi'\right) \label{eq:F}\,,\\ 
\label{eq:dBoverB} \frac{\mathcal{B}'}{\mathcal{B}}\,&=\frac{20 u \phi ''-9 u \phi '^2+24 \phi '}{2 \left(5 u \phi
   '+12\right)}\,,
\end{align}
where again a prime denotes a differentiation with respect to $u$. For the dilaton one obtains a third order differential equation that is shown in \eqref{eq:phiDGLMT} in appendix \ref{app:equ}. Again one has to specify the boundary conditions. In particular $\phi(0) = 0$ and $\mathcal{F}(0) = \mathcal{H}(0) = \mathcal{B}(0) = 1$ (see \cite{Mateos:2011ix} for the rest). 

The latter construction of an anisotropic model for the quark gluon plasma has been used frequently since its presentation in \cite{Mateos:2011tv}. Interesting observables in this model are for example the binding energy and screening distance of a $q\bar q$-pair \cite{Giataganas:2012zy,Rebhan:2012bw,Chernicoff:2012bu,Chakraborty:2012dt}, jet quenching \cite{Giataganas:2012zy}, the drag force on heavy quarks \cite{Giataganas:2012zy,Chernicoff:2012iq}, thermal photon production \cite{Patino:2012py}, and Brownian motion/diffusion \cite{Giataganas:2013hwa,Giataganas:2013zaa,Chakrabortty:2013kra}. Another interesting feature of this construction is the violation of the previously predicted lower bound for the ratio of the shear viscosity over the entropy density $\eta/s = 1/4\pi$. 

Our idea is to extend the present construction by combining it with the non-conformal deformation of section \ref{sec:deform1}. This is shown in the following section.  

\subsection{Two-Parameter Combined Deformation}
\label{sec:twoParam}
\begin{figure}[ht]\centering
 \includegraphics[scale=1]{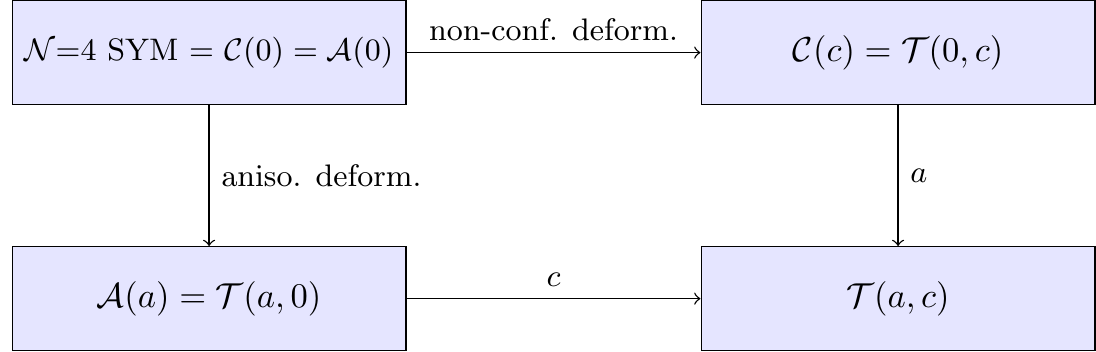}
 \caption{Scheme of the two-parameter deformation we seek. $\mathcal{A}(a)$ symbolizes the anisotropic deformation with deformation parameter $a$ and $\mathcal{C}(c)$ the isotropic non-conformal deformation of SYM theory with parameter $c$. $\mathcal{T}(a,c)$ symbolizes the model we want to construct. }
 \label{fig:scheme}
\end{figure}

So far we have introduced models that describe isotropic non-conformal deformations in section \ref{sec:deform1} and a second model that describes an anisotropic deformation in \ref{sec:MT}. The deformations each can be controlled by a single parameter where we can always obtain \SYM~  continuously for vanishing deformation parameter. The goal of the present section is to construct a model of combined non-conformal and anisotropic deformation. We seek a two-parameter deformation where one parameter -- call it $c$ -- controls the non-conformal deformation and the other parameter -- call it $a$ -- controls the anisotropic deformation. For vanishing $c$ we want to obtain continuously the anisotropic model of section \ref{sec:MT}, whereas for vanishing $a$ we want to get a purely non-conformal deformation. The general idea of the two-parameter deformation we seek is shown in figure \ref{fig:scheme}.

We want a model that is given by a solution to equations of motion of a (super)gravity action in 5D. Therefore we first have to introduce an action that combines the different models. For anisotropy we need the axion-dilaton $(\phi,\chi)$ and for the non-conformal deformation we need an additional scalar $\tau$ and a potential $V(\tau)$. At this stage, one might wonder if it is not easier to simply introduce a potential for $\phi$ or $\chi$.  However, generic potentials in the 5D gravity-axion-dilaton-action lead to inconsistencies because the system of differential equations derived from the action is over-determined. We will talk more about this in what follows, too. 

In string inspired models, all fields couple to the dilaton. The coupling is given by a factor $e^{b\phi}$, where $b = \text{const.}$ controls the strength of the coupling. Thus, we first assume that the additional scalar is coupled to the dilaton, too. The action in five dimensions can then be written as
\begin{equation} 
 S = \frac{1}{16 \pi G}\int_\mathcal{M} d^5x\,\sqrt{-g}\, \left(\mathcal{R} +12 -\frac{1}{2} (\partial\phi)^2 -\frac{1}{2}e^{2\phi}(\partial\chi)^2 - \frac{1}{2}e^{b\phi}\left( (\partial\tau)^2 +2 V(\tau)\right)\right).
\end{equation}

\noindent
We choose the same Ansatz for the metric in Einstein frame as in section \ref{sec:MT}
\begin{equation}
 ds^2 = \frac{e^{-\frac{1}{2}\phi}}{u^2}\left(-\mathcal{FB}~ dt^2 +dx^2 + dy^2 + \mathcal{H} dz^2 +\frac{du^2}{\mathcal{F}}\right)\,,
\end{equation}
where all the metric functions and the dilaton $\phi$ and the scalar $\tau$ depend only on the holographic coordinate $u$. As in section \ref{sec:MT}, the axion and the metric function $\mathcal{H}$ can be chosen to be
\begin{equation}
 \chi(z) = a\,z~~~\text{and}~~~ \mathcal{H}(u) = e^{-\phi(u)}.
\end{equation}

\noindent
Let us now summarize all the unknown functions: We have two metric functions $\mathcal{F}(u)$ and $\mathcal{B}(u)$, the dilaton $\phi(u)$, and the scalar $\tau(u)$. They are determined by solving the following equations: five Einstein equations, the equation of motion for the dilaton, the equation of motion for the axion, and the equation of motion for the scalar $\tau$. Due to rotational symmetry in the $xy$-plane, the $xx$- and the $yy$-Einstein equations are equivalent. The equation of motion for the axion is solved by the Ansatz $\chi(z) = a\,z$. Therefore we are left with six coupled differential equations for three unspecified objects. The system is generically over-determined and many possible potentials lead to a system that cannot be solved. It turns out that for the potential $V(\tau) = -2\tau^2 - c \tau^4$ as in section \ref{sec:deform1} only a vanishing coupling of the scalar to the dilaton, i.e. $b=0$, lead to consistent solutions. Thus, we choose the action to be
\begin{equation}
 S = \frac{1}{16 \pi G}\int_\mathcal{M} d^5x\,\sqrt{-g}\, \left(\mathcal{R} +12 -\frac{1}{2} (\partial\phi)^2 -\frac{1}{2}e^{2\phi}(\partial\chi)^2 - \frac{1}{2}(\partial\tau)^2 - V(\tau)\right).\label{eq:wrightaction}
\end{equation}

\noindent
With this action, the remaining metric functions can be given in terms of the dilaton $\phi$ and the scalar $\tau$ as
\begin{align}
 \mathcal{\frac{B'}{B}} &= \frac{-8 u \left(\tau '\right)^2+24 \phi '-9 u \left(\phi '\right)^2+20 u \phi ''}{24+10 u \phi '}\,,\\
 \mathcal{F} &= \frac{e^{-\phi /2} \left(4 \left(12+2
\tau ^2+c \tau ^4\right) \phi '+3 a^2 e^{7 \phi /2} u \left(4+u \phi '\right)\right)}{12 \left(\phi '+u \phi ''\right)}\,.
\end{align}

\noindent
The dilaton and the additional scalar can be determined by solving two coupled differential equations of third order $\phi$ and second order in $\tau$. They are explicitly given in \eqref{eq:combDGL1} and \eqref{eq:combDGL2} in appendix \ref{app:equ}. One can solve these equations numerically by choosing the same boundary conditions for $\tau$ as in the model of section \ref{sec:deform1} and for $\phi$ as in the anisotropic model of section \ref{sec:MT}. 

The construction above is a consistent\footnote{By consistent we mean that it solves all the equations of motion of a gravity action.} two-parameter deformation of AdS$_5$ space, where one parameter originates from an anisotropic deformation and the other from a non-conformal deformation. Within numerical accuracy, we reproduce the known models of section \ref{sec:deform1} for $a\rightarrow0$ and of section \ref{sec:MT} for $c\rightarrow0$ and, hence, our model has in fact the structure of figure \ref{fig:scheme}. By solving the coupled differential equations \eqref{eq:combDGL1} and \eqref{eq:combDGL2} we can compute the metric background of our model for many combinations of $(a,c,u_\text{h})$.\footnote{In the following section we will introduce the temperature in this model. Then we can replace $u_\text{h}$ by the more physical parameter $T$.} 

In the following we will first investigate the thermodynamics and in particular show how temperature and entropy density behave in the different deformations. Thereby we will replace the (from the plasma's viewpoint) unphysical parameter $u_h$ by the temperature/entropy density of the plasma. 

\subsection{Temperature and entropy density in the different models} 

So far the geometries of all the models we presented depend on $u_\text{h}$ -- the specific position of a black brane horizon. The position $u_\text{h}$ itself has no physical meaning. However, for a black brane one can define the well-known Hawking temperature \cite{Hawking:1973bh} and the entropy density by the Bekenstein-Hawking formula. As shown in appendix \ref{app:TandS} for any of our models the temperature can be given in terms of the metric functions as
\begin{equation}
 T = \frac{\mathcal{F}'(u_\text{h})\sqrt{\mathcal{B}(u_\text{h})}}{4\pi},
\end{equation}
where a prime again denotes a derivative with respect to $u$. We can see that the temperature explicitly depends on the position of the horizon but also implicitly depends on any deformation parameter through the metric functions. The entropy density is given by the Bekenstein-Hawking formula \cite{Bekenstein:1973ur} as 
\begin{equation}
 s = \left\{\begin{matrix}
             \frac{N_c^2}{2\pi} \,\frac{1}{u_\text{h}^3}~~~~~ \text{for the isotropic model}\\
             \frac{N_c^2}{2\pi} \,\frac{e^{-\frac{5}{4}\phi(u_\text{h})}}{u_\text{h}^3} ~~~ \text{for the anisotropic models}
            \end{matrix}\right.\,,
\end{equation}
where $N_c \propto 1/\sqrt{G}$ is the number of colors in the corresponding gauge theory that we assume to be the same and constant in all our models. One can see that for the isotropic non-conformal deformation the entropy density does not depend on the deformation parameters whereas for the anisotropic deformations it does through the dilaton $\phi$.

\begin{figure}[!ht]\centering
 \includegraphics[width=0.49\textwidth]{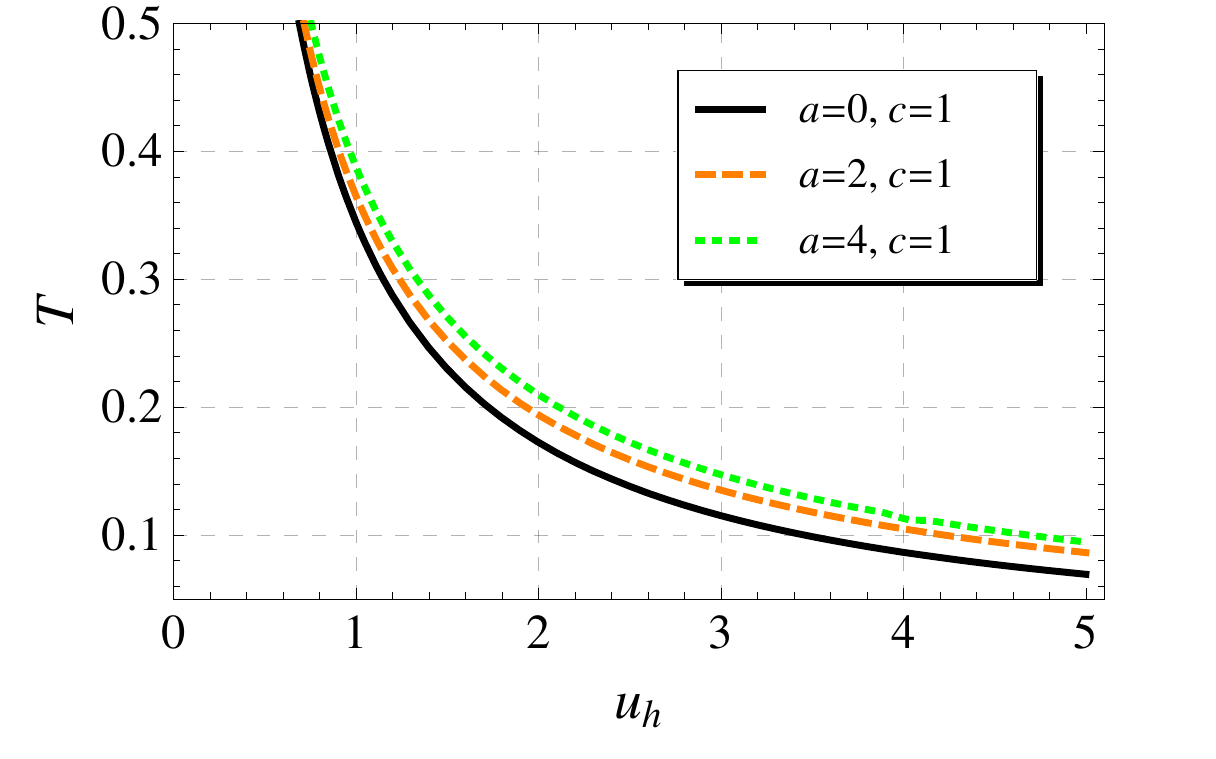}
 \includegraphics[width=0.49\textwidth]{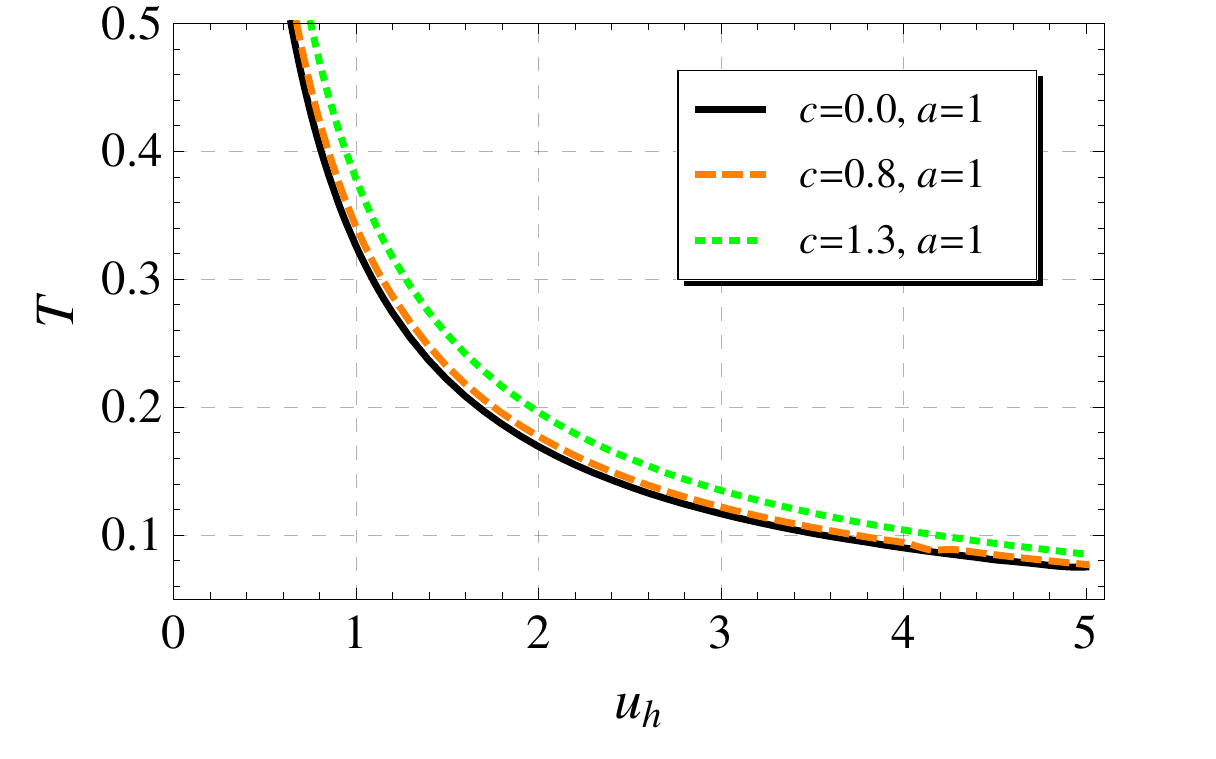}
 \includegraphics[width=0.49\textwidth]{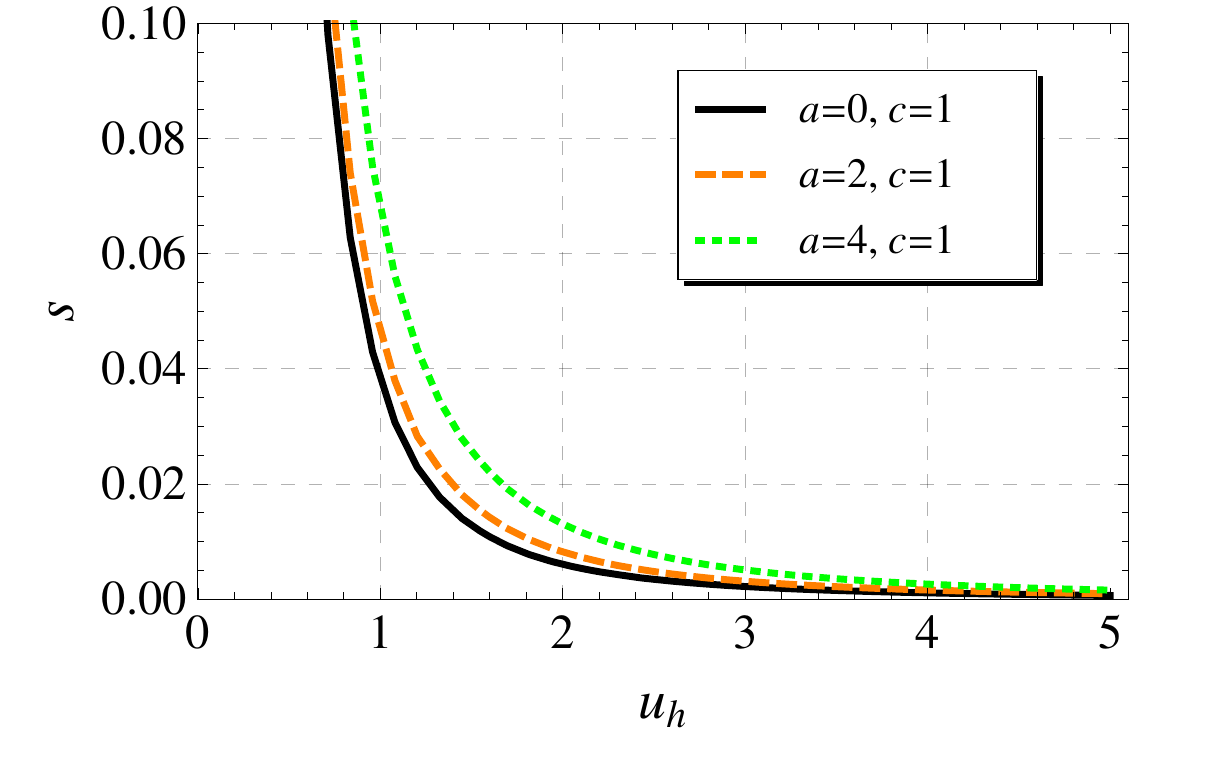}
 \includegraphics[width=0.49\textwidth]{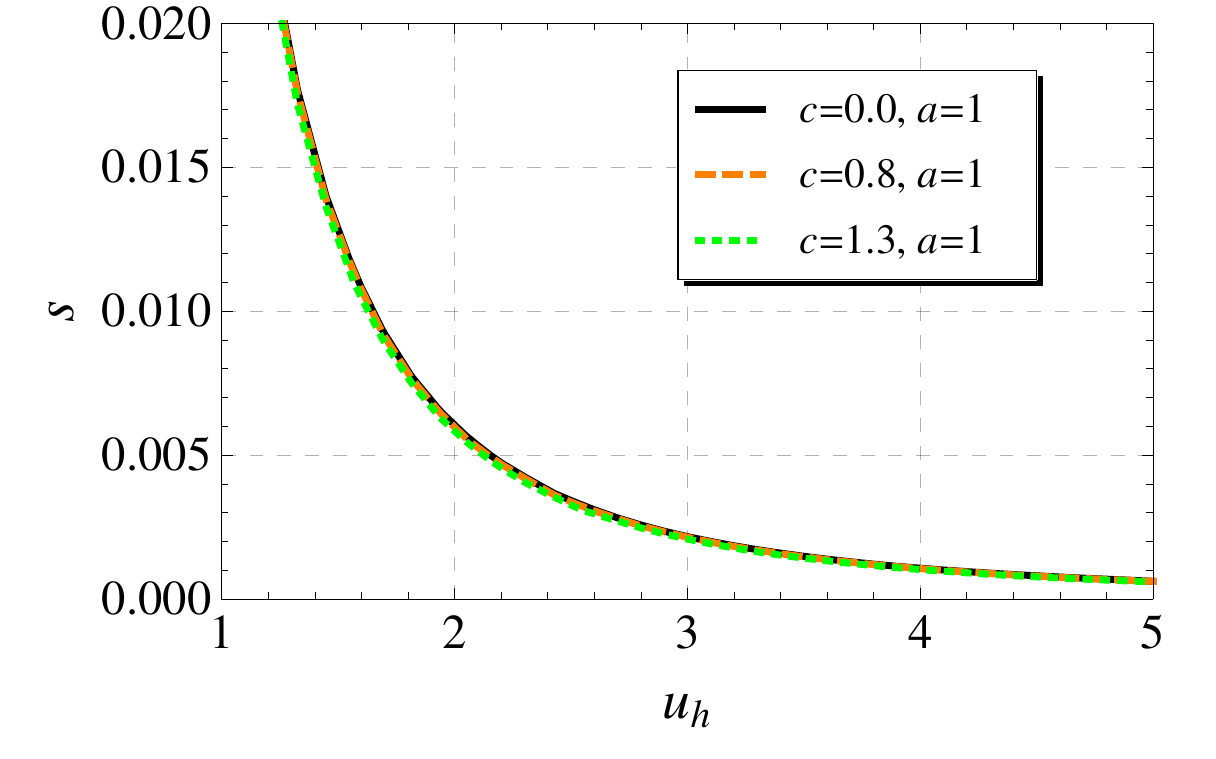}
 \caption{Visualization of the temperature $T$ and the entropy density $s$ as a function of $u_\text{h}$ for different values of the parameters $a$ and $c$. The shapes of the curves always look alike, but the temperature and the entropy density increase for growing $a$. For growing $c$ only the temperature shows a significant change. The numerics indicates that the maps $u_\text{h} \mapsto T$ and $u_\text{h} \mapsto s$ are one-to-one and onto for all values of the parameters.}
 \label{fig:TundsxuhComb}
\end{figure}

In the following we want to consider the temperature and the entropy density in the two parameter model of section \ref{sec:twoParam}. Since the model and in particular the metric functions reduce to the corresponding one-parameter isotropic model for $a\rightarrow0$ and to the one-parameter anisotropic model for $c\rightarrow0$, both the temperature and the entropy density become the respective quantities in these limits, too. In figure \ref{fig:TundsxuhComb} we visualize $T$ and $s$ as  functions of the horizon position. 

In contrast to the horizons position $u_\text{h}$, both the temperature and the entropy density are \text{physical} quantities -- i.e. in principle measurable in the gauge theory. In what follows we want to derive other observables and in particular analyze their behavior under the deformations at a fixed temperature (or entropy) rather than a fixed horizon position $u_\text{h}$. 

Another interesting physical quantity that now can be derived easily is the specific heat. It contains important information about the thermodynamics of the plasma and can be derived via
\begin{equation}
 C = T \,\frac{\partial s}{\partial T}\,.
\end{equation}

\begin{figure}
\centering
\includegraphics[scale=.85]{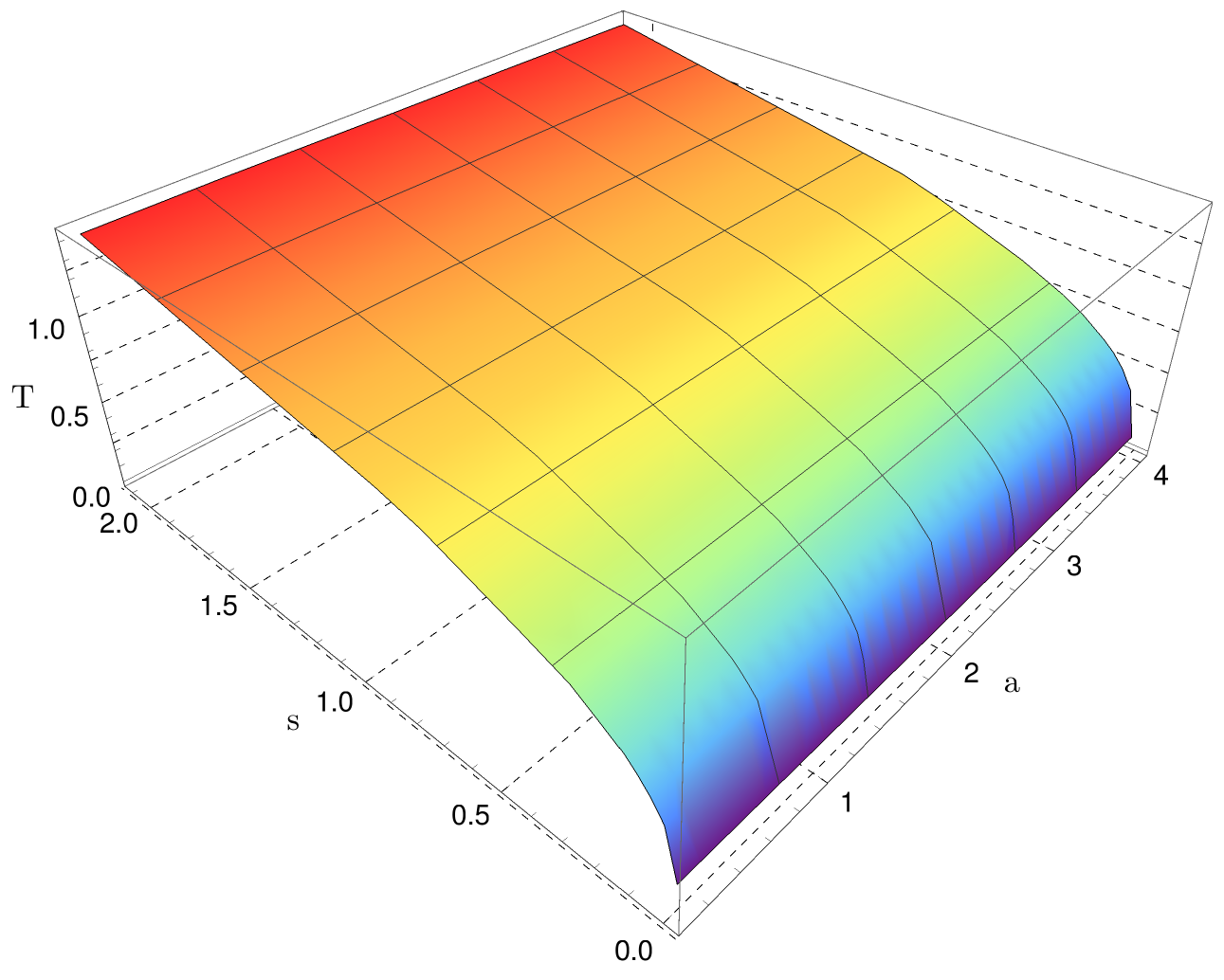}
\label{fig:Txsxa}
\caption{The temperature of the anisotropic model as a function of the entropy density and the anisotropy parameter at fixed non-conformal deformation (c=1). It is visible that $T$ does not change at all with $a$, which in particular shows that the specific heat $C$ of the plasma does not change under the anisotropic deformation.}
\end{figure}

\noindent
Since the previous maps $u_\text{h}\mapsto T,s$ are bijective it is allowed to define a map $T\mapsto s$. The (numerical) solutions is
\begin{equation}
 s(T) = \tilde C(c)\, T^3\,
\end{equation}
where $\tilde C(c)$ does not depend on $a$ but increases for larger non-conformal deformation parameter $c$. As a consequence, the specific heat $C = 3\,\tilde C(c) \,T^3$ itself does not depend on the \textit{anisotropy parameter} $a$! This is in particular visible in figure \ref{fig:Txsxa}, where we show the dependency of $T$ on $s$ and $a$ at fixed $c=1$. It is easy to see that $T(s)$ is the same for all considered values of $a$. 

\section{Heavy Quarks in the Deformed Model}\label{sec:observables}

\subsection{Screening and Binding Energy}

We now want to investigate the binding energy and the screening distance of a $q\bar q$-pair of a heavy quark and a heavy antiquark in a bound state in the gauge/gravity set-up. The screening distance is defined as the maximal distance where a bound state in the hot plasma is still possible. On the gauge theory side the object of choice to derive the latter two quantities are Wilson loops $\mathcal{W}$ \cite{PhysRevD.10.2445}. For a rectangular Wilson loop that is largely extended in the time direction, its expectation value can be written as
\begin{equation}
 \langle\mathcal{W}\rangle = \exp(-i\mathcal{T} E(L))\,,
\end{equation}
where $\mathcal{T}\gg 1$ is the extension of the Wilson loop in the time direction, $L$ is its extension in the spatial direction and can be interpreted as the distance of a $q\bar q$-pair, and  $E(L)$ is the energy of the $q\bar q$-pair. 

The dual description for Wilson loops can be given in term of string worldsheets \cite{Liu:2006he}. When one interprets the Wilson loop as the path traversed by a quark, then the dual description to quarks are open strings attached to D-branes sitting at some holographic position that is inversely proportional to their mass. Thus for infinitely heavy -- non-dynamical -- quarks the boundary of the string worldsheet lies at the boundary of the gravity description \cite{Rey:1998ik}. This is the limit we want to work in. One now identifies the partition function of the Wilson loop with the partition function of the string worldsheet. In the limit of strong coupling, the string partition function simplifies to the exponential of the on-shell string/Nambu-Goto action $S_\text{NG}$, so that we can write
\begin{equation}
 \langle\mathcal{W}\rangle = \exp(iS_\text{NG})\,.
\end{equation}

\noindent
Let us now assume a non-dynamical quark-antiquark pair in the anisotropic plasma whose dual description is given by geometries derived in section \ref{sec:twoParam}. To derive the binding energy and the screening distance we have to consider a rectangular Wilson loop with large $\mathcal{T}$. The dual object is a string 'hanging' from the boundary towards the horizon into the geometry specified by the metric
\begin{equation}
 ds^2 = \frac{1}{u^2} \left(-\mathcal{FB}~ dt^2 +dx^2 + dy^2 + e^{-\phi}\, dz^2 +\frac{du^2}{\mathcal{F}}\right)\,.
\end{equation}

\begin{figure}[t]
 \centering
 \includegraphics[scale=1]{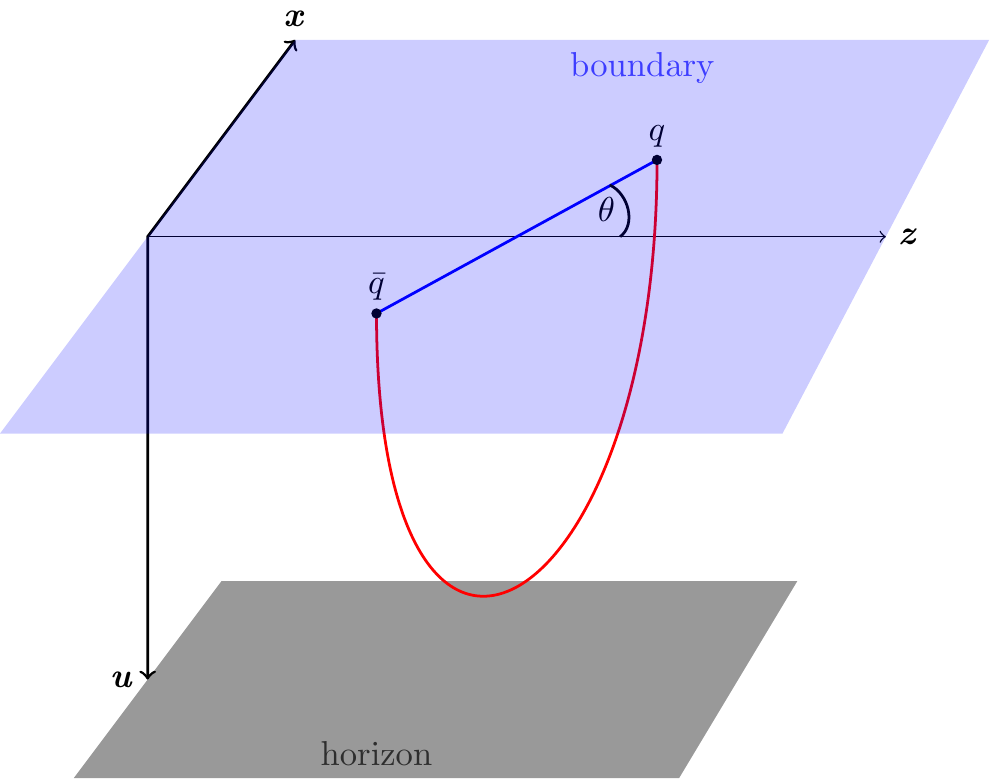}
 \caption{Schematic picture of the string configuration that describes a $q \bar q$-pair in the anisotropic theory. The connecting line has the angle $\theta$ to the anisotropic direction $z$.}
 \label{fig:pairaniso} 
\end{figure}

\noindent
We expect that observables like the screening distance and the binding energy of a $q \bar q$-pair depend on the orientation relative to the anisotropic direction. To check this expectation we introduce the angle between the $z$-axis and the direction of $q\bar q$-pair as an additional parameter. In figure \ref{fig:pairaniso} we show a $q \bar q$-pair in its dual description and its angle with the anisotropic direction $z$. To parametrize the shown configuration we choose 
\begin{equation}
\begin{split}
  t &= \iota\,,~~~ x = \sigma\,,~~~z = z(\sigma)\,,\\
  u &= u(\sigma)\,, ~~~\text{and}~~~ y = \text{const. }.
\end{split}
\end{equation}

\noindent
The induced worldsheet metric is then given by
\begin{equation}
 ds^2_\text{\tiny WS} = \frac{1}{u^2} \left(-\mathcal{FB} \,d\iota^2 + \left(1 + e^{-\phi} (z')^2 + \frac{(u'^2)}{\mathcal{F}}\right) d\sigma^2 \right),
\end{equation}
where a prime denotes a derivative with respect to $\sigma = x$. The Nambu--Goto action then reads
\begin{align}
 S &= -\frac{1}{2 \pi \alpha'} \int d\iota\,d\sigma\,\sqrt{-\det g_\text{\tiny WS}} \nonumber\\
   &= -\frac{\mathcal{T}}{2\pi\alpha'} \int d\sigma\, \frac{1}{u^2} \sqrt{\mathcal{FB}\left(1+e^{-\phi}(z')^2 + \frac{(u')^2}{\mathcal{F}}\right)} \label{eq:strinactionComb}\\
   &\equiv -\frac{\mathcal{T}}{2\pi\alpha'} \int d\sigma \, \mathcal{L}\,, \nonumber
\end{align}
where we define the Lagrangian $\mathcal{L} =  u^{-2} \sqrt{\mathcal{FB}\left(1+e^{-\phi}(z')^2 + \frac{(u')^2}{\mathcal{F}}\right)}$\,. We can now derive the equations of motion for this action. 
First we see that $\mathcal{L}$ has no $\sigma$-dependence, so that the corresponding Hamiltonian $H$ is a constant of motion:
\begin{align}
 c_1 = H &= \mathcal{L} - z'\frac{\partial \mathcal{L}}{\partial z'} - u' \frac{\partial \mathcal{L}}{\partial u'}\\
 \Rightarrow ~~ \mathcal{L} \,c_1 &= \frac{\mathcal{FB}}{u^4}\,.
\end{align}
Next we see that $\mathcal{L}$ does not depend on $z$ but only on $z'$. So the Euler--Lagrange equation for $z$ simplifies to
\begin{align}
 c_2 = \frac{\partial\mathcal{L}}{\partial z'} = e^{-\phi}\,\frac{\mathcal{FB}}{u^4}\,\frac{z'}{\mathcal{L}}~~\Rightarrow~~ \mathcal{L} \, c_2 = e^{-\phi} \frac{\mathcal{FB}\,z'}{u^4}\,.
\end{align}
Combining both equations leads to the two rather simple equations of motion for $u$ and $z$:
\begin{align}
 z' &= \frac{c_2}{c_1} e^\phi\,,\label{eq:zeomcomb}\\
 u'^2 &= \frac{\mathcal{F}}{c_1^2}\left( \frac{\mathcal{FB}}{u^4} -c_1^2 - c_2^2\,e^\phi \right)\,. \label{eq:ueomcomb}
\end{align}

\noindent
The string configuration as shown in figure \ref{fig:pairaniso} has a maximum for $u(\sigma)$ at $\sigma = 0$. There the string turns around. This gives rise to two constraints: 
\begin{equation}\label{eq:turningPoint}
 u(0) \equiv u_\text{t}\,,~~~~ u'(0) = 0\,,
\end{equation}
where $u_t$ is called the turning point. Thus, at the turning point we obtain
\begin{equation}\label{eq:ellipse}
 c_1^2 + c_2^2\,e^{\phi_\text{t}} = \frac{\mathcal{F}_\text{t}\mathcal{B}_\text{t}}{u_\text{t}^4}\,,
\end{equation}
where a subscript "$t$" denotes a function evaluated at $u_\text{t}$. We can now parametrize the two constants $c_1$ and $c_2$ by the turning point $u_\text{t}$ and an angle $\varphi$. The latter appears when we interpret \eqref{eq:ellipse} as the defining equation of an ellipse in the $(c_1,c_2)$-plane. We then can write
\begin{align}
 c_1^2 &= \frac{\mathcal{F}_\text{t}\mathcal{B}_\text{t}}{u_\text{t}^4} \sin^2\varphi\,, \\
 c_2^2 &= \frac{\mathcal{F}_\text{t}\mathcal{B}_\text{t}}{u_\text{t}^4}\, e^{-\phi_\text{t}} \cos^2\varphi\,.
\end{align}

\subsubsection{Screening distance}

We use the latter results to calculate the distance $L$ and the angle $\theta$ to the anisotropic direction of the $q\bar q$-pair:
\begin{align}
 \frac{L}{2}\,\sin\theta &= \int_0^{\frac{L}{2}\sin\theta} dx = \int_0^{u_\text{t}} \frac{du}{u'} = \int_0^{u_\text{t}} du\,\frac{c_1}{\sqrt{\mathcal{F}\left(\frac{\mathcal{FB}}{u^4} - c_1^2 - c_2^2\,e^\phi\right)}} \nonumber\\
 &\equiv  \frac{\sqrt{\mathcal{F}_\text{t}\mathcal{B}_\text{t}} \,\sin\varphi}{u_\text{t}^2} \int_0^{u_\text{t}} du ~\mathcal{I}(u,u_\text{t},\varphi)\,,\label{eq:getL1}\\[2em]
 \frac{L}{2}\,\cos\theta &= \int_0^{\frac{L}{2}\cos\theta} dz = \int_0^{u_\text{t}}du \frac{z'}{u'} = \frac{\sqrt{\mathcal{F}_\text{t}\mathcal{B}_\text{t}} \,\cos\varphi}{u_\text{t}^2~e^{\frac{1}{2}\phi_\text{t}}} \int_0^{u_\text{t}} du~ e^\phi ~\mathcal{I}(u,u_\text{t},\varphi)\label{eq:getL2}\\[2em]
 \Rightarrow ~~~\tan\theta &= \tan\varphi ~ \frac{\int_0^{u_\text{t}}du~\mathcal{I}(u,u_\text{t},\varphi)}{\int_0^{u_\text{t}}du~e^{\phi-\phi_t/2}~\mathcal{I}(u,u_\text{t},\varphi)}\,, \label{eq:gettheta}
\end{align}
where we define $\mathcal{I}(u,u_\text{t},\phi) \equiv \left[\mathcal{F}\left(\frac{\mathcal{FB}}{u^4} -\frac{\mathcal{F}_\text{t}\mathcal{B}_\text{t}}{u_\text{t}^4}-\left(e^{\phi-\phi_\text{t}}-1\right)\frac{\mathcal{F}_\text{t}\mathcal{B}_\text{t}}{u_\text{t}^4}\cos^2\varphi\right)\right]^{-\frac{1}{2}}$. For every given pair $(u_\text{t},\varphi)$ we first evaluate \eqref{eq:gettheta} to obtain the angle $\theta(u_\text{t},\varphi)$ and then integrate \eqref{eq:getL1} or \eqref{eq:getL2} to get the distance $L(u_\text{t},\varphi)$. So our procedure gives a map $(u_\text{t},\varphi) \mapsto (L,\theta) \,.$

For our anisotropic two-parameter model of section \ref{sec:twoParam} we have to evaluate the integrals numerically. A first observation is that the difference of the angles $\theta$ and $\varphi$ turns out to be small. The difference is always computed to be in the interval $(-5^\circ,5^\circ)$, usually it is much smaller. If the $q\bar q$-pair lies parallel ($\theta=0$) or perpendicular ($\theta = \pi/2$) to the anisotropic direction the difference actually vanishes. To exactly get $L(u_t,\theta)$ for a given $\theta\neq0,\pi/2$ we have to fine-tune $\varphi$. 

We can use the procedure explained above to derive the distance of a $q\bar q$-pair for every combination $(a,c,u_\text{t},\theta)$ at fixed temperature $T$. For this we first have to solve the equations of motion for the metric functions and the dilaton for given $a$, $c$ and $T$. Then we have to evaluate the integrals \eqref{eq:getL1}--\eqref{eq:gettheta}. In this way we obtain $L(a,c,u_\text{t},\theta)$ whose maximum for fixed $a$, $c$ and $\theta$ gives the screening distance $L_s(a,c,\theta)$. 

\begin{figure}[!ht]\centering
 \includegraphics[width=.7\textwidth]{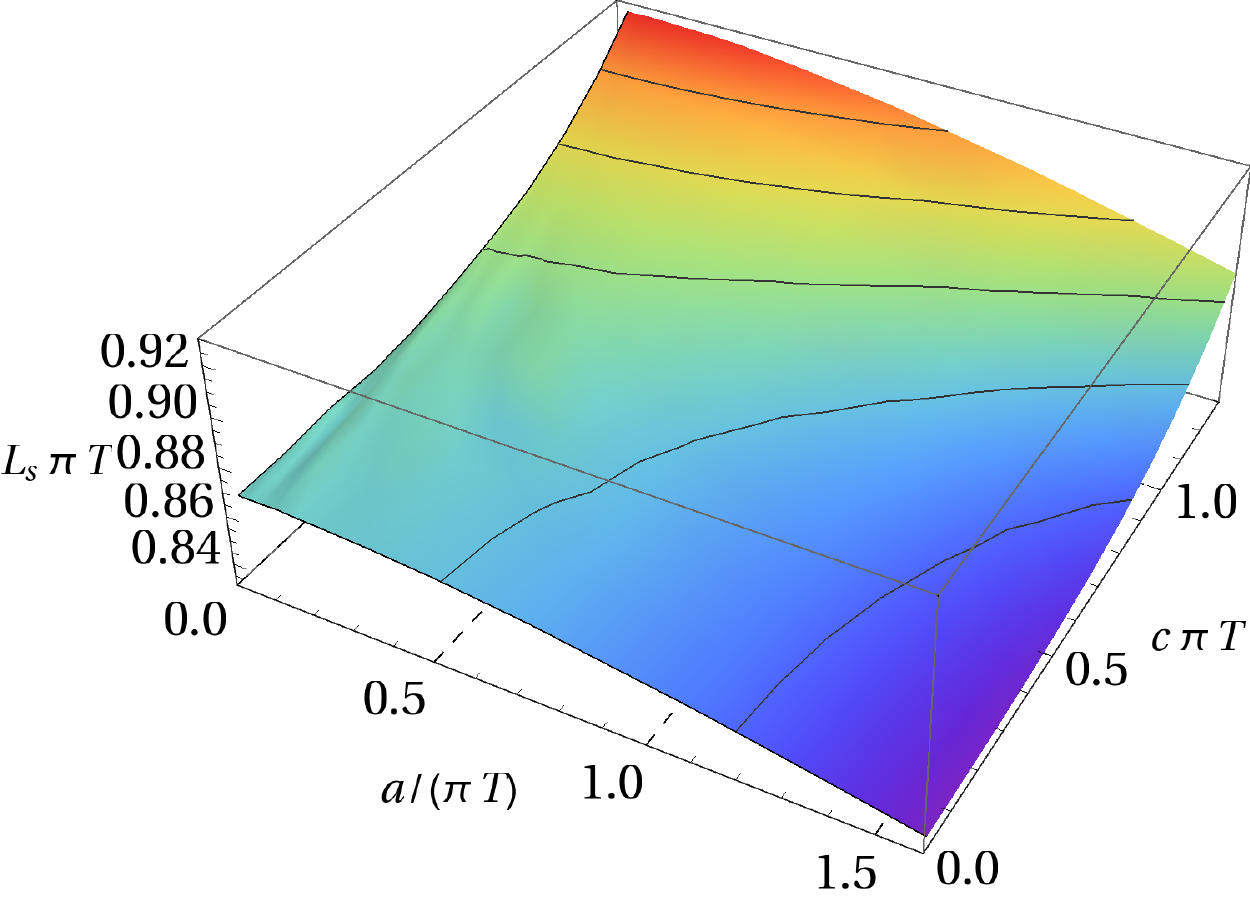}
 \caption{The screening distance as a function of the two parameters $c$ and $a$ of our model for a $q\bar q$-pair lying parallel to the anisotropic direction. $L_s$ increases for growing $c$ and decreases for growing $a$. }
 \label{fig:screeningcomb1} 
\end{figure}

\begin{figure}[t]\centering
 \includegraphics[width=.45\textwidth]{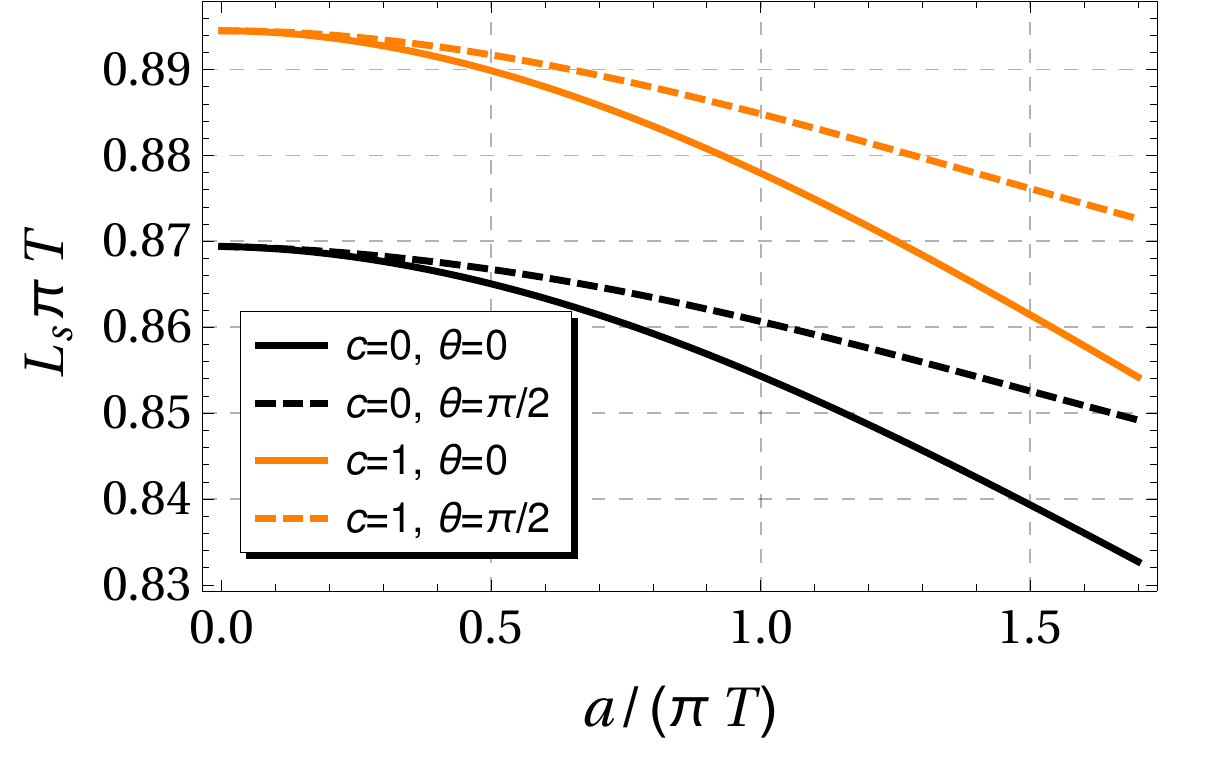}
 \includegraphics[width=.45\textwidth]{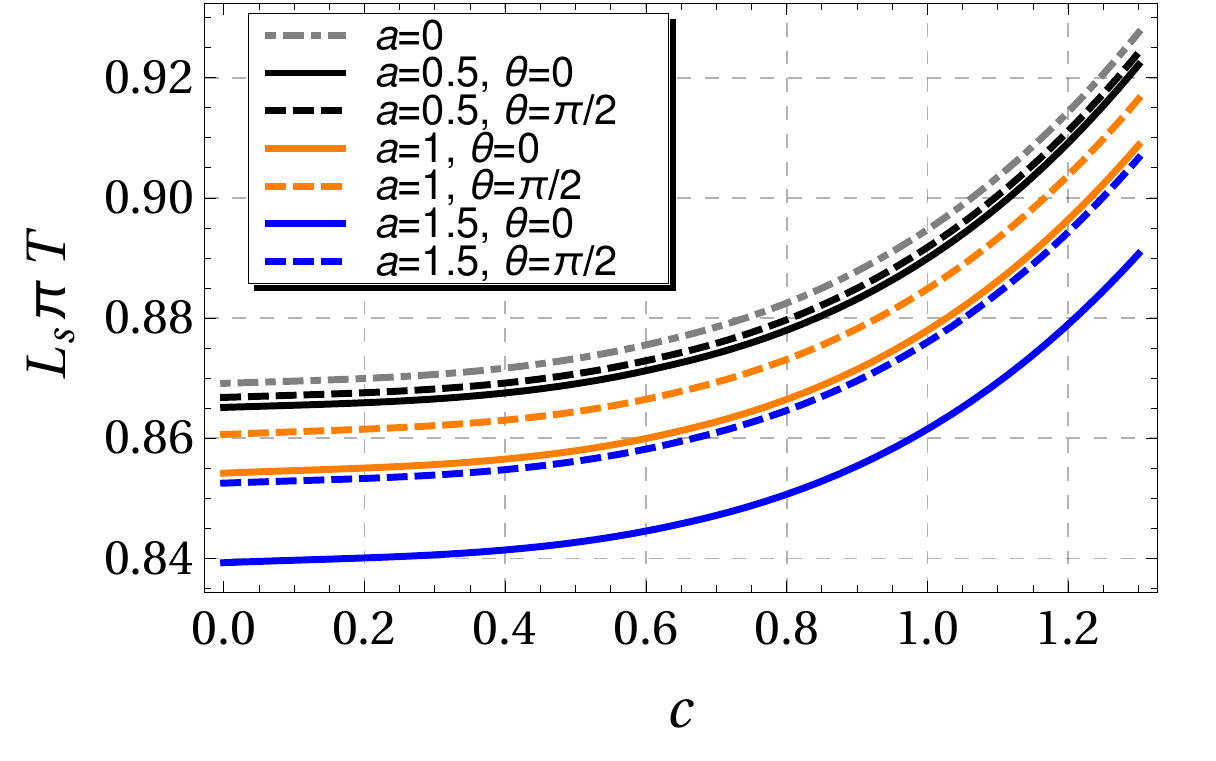}
 \caption{Evolution of the screening distance for growing parameter $a$ (left) and $c$ (right) for different values of the respective other deformation parameter, parallel and perpendicular to the anisotropic direction. $L_s$ is always decreasing in $a$ and increasing in $c$, so that the screening distance always behaves in both parameters as in the corresponding one-parameter deformation. }
 \label{fig:screeningcomb2} 
\end{figure}

\begin{figure}[ht]\centering
 \includegraphics[width=.45\textwidth]{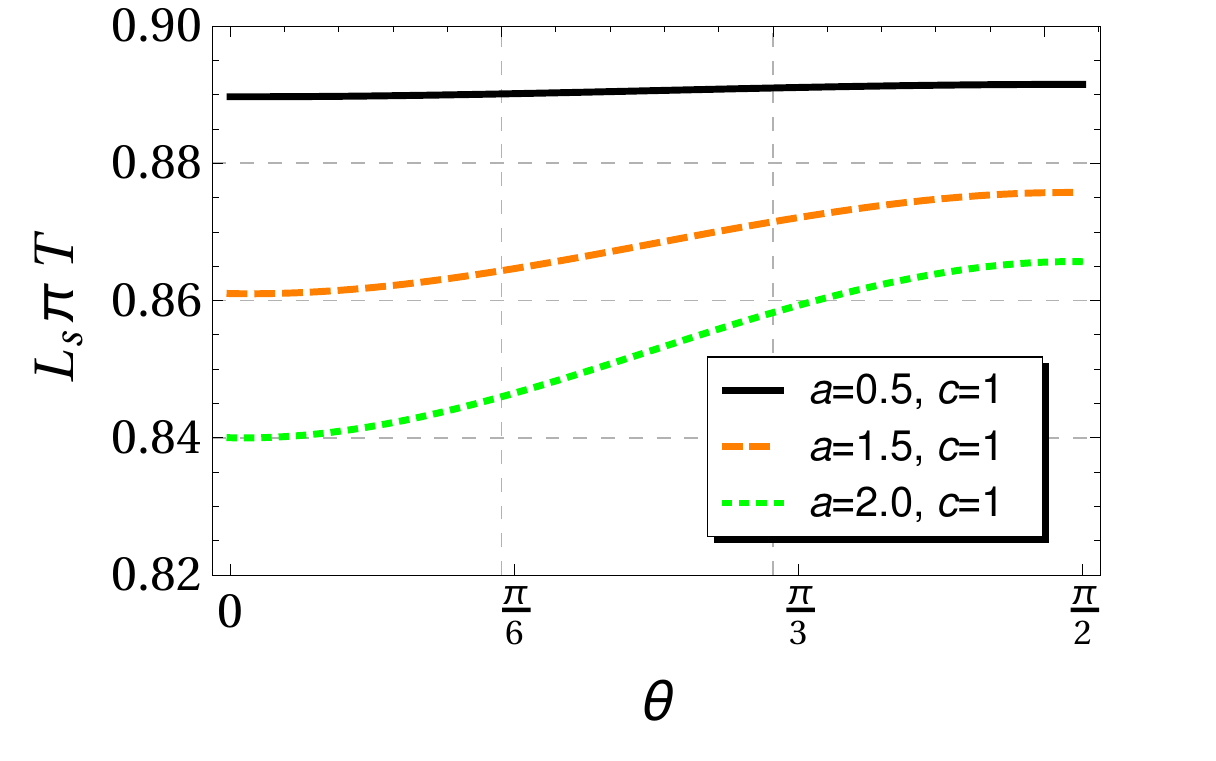}
 \includegraphics[width=.45\textwidth]{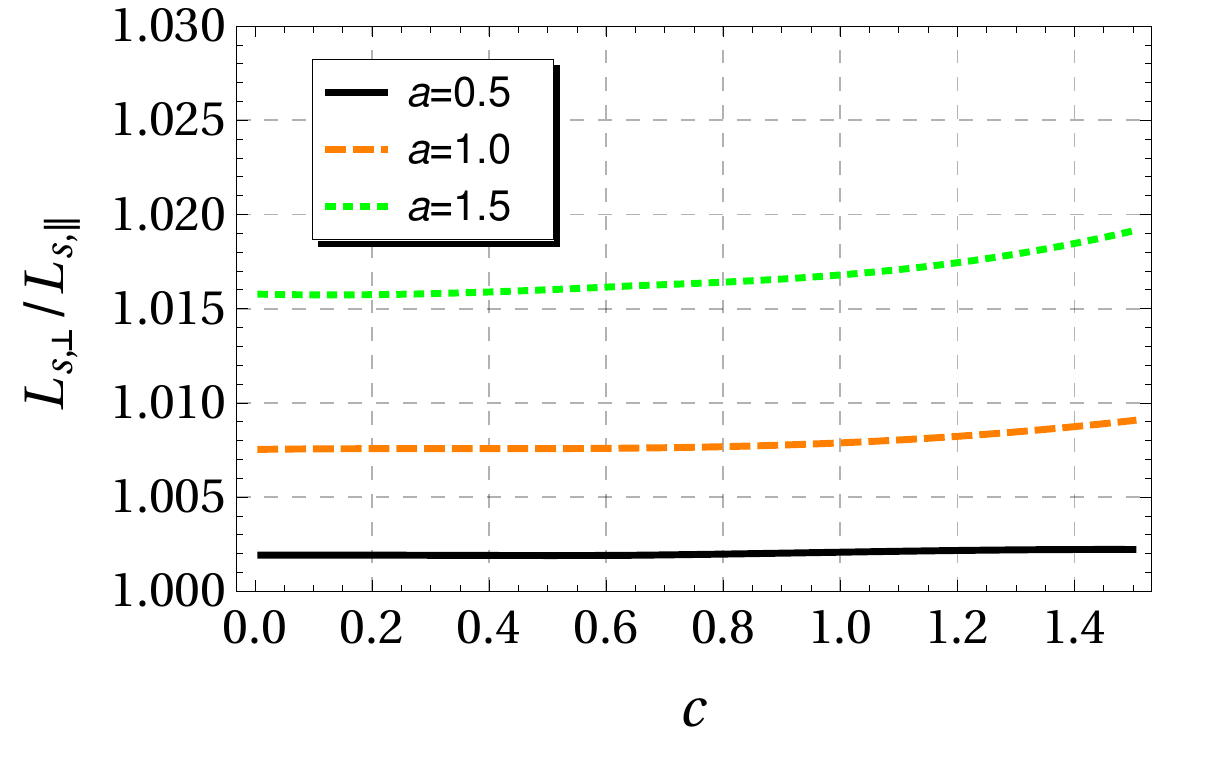}
 \caption{The screening distance as a function of the angle between the anisotropic direction and the $q \bar q$-pair (left) and the ratio of the screening distance parallel and perpendicular as a function of $c$ (right). The left figure shows that the larger $a$ the large the angle affects the screening distance. The right figure shows that $c$ only marginally affects the dependence of $L_s$ on the orientation.}
 \label{fig:screeningcomb5} 
\end{figure}

Results for the screening distance are illustrated in the figures \ref{fig:screeningcomb1}--\ref{fig:screeningcomb5}. The numerical results show that the screening distance increases for larger $c$ and decreases for larger $a$. In a more mathematical language this is
\begin{align}
 \forall_{(c_0,a_0)} ~:~\partial_c L_s\vert_{(c_0,a_0)} > 0 \,,\\
 \forall_{(c_0,a_0)} ~:~\partial_a L_s\vert_{(c_0,a_0)} < 0 \,. 
\end{align}

\noindent
The latter is in particular true in the two one-parameter models where $c=0$ or $a=0$, respectively, and is in no conflict with earlier results on the screening distance \cite{Ewerz:2010du,Giataganas:2012zy}. The numerical results show that the quantitative effect of the two parameters on the screening distance is always the same. This will be important for our conclusion and final interpretation of the two parameters.    

In figure \ref{fig:screeningcomb5}, we show how the screening distance is influenced by the orientation of the $q\bar q$-pair relative to the anisotropic direction and how $a$ and $c$ affect this influence. It is evident that $L_s$ is affected most when the pair is oriented parallel to the anisotropic direction. This is what one would expect because of the stronger deformation of the geometry in this direction. We can also see that the larger $a$ the stronger is the effect of the orientation on the screening distance, which is expected, too, since the parameter $a$ is designed to control the anisotropy. We want to mention that $c$ influences the effect of the orientation of the screening distance for fixed $a$, too. On the right of figure \ref{fig:screeningcomb5}, we can see that $c$ affect $L_{s\perp}/L_{s\parallel}$ and thus the influence of the orientation for $a\neq0$. However, the effect of the deformation parameter $c$ is rather small compared the effect of $a$. 

\subsubsection{Binding energy}

Let us now return to the calculations. We want to derive a formula for the binding energy, so we have to return to \eqref{eq:strinactionComb} and use the equations of motion \eqref{eq:zeomcomb} and \eqref{eq:ueomcomb} to obtain the on-shell string action
\begin{equation}
\begin{split}
 S &= -\frac{\mathcal{T}}{\pi\alpha'} \int_0^{u_\t} \frac{du}{u'}\, \frac{\mathcal{FB}}{c_1u^4} = -\frac{\mathcal{T}}{\pi\alpha'}\int_0^{u_\t} du\,\frac{\mathcal{FB}}{u^4\sqrt{\mathcal{F}\left(\frac{\mathcal{FB}}{u^4} - c_1^2 - c_2^2\,e^\phi\right)}}\\
   &= -\frac{\mathcal{T}}{\pi\alpha'} \int_0^{u_\t} du\,\frac{\mathcal{BF}}{u^4}\,\left[\,\mathcal{F}\left(\frac{\mathcal{FB}}{u^4} -\frac{\mathcal{F}_\t\mathcal{B}_\t}{u_\t^4}-\left(e^{\phi-\phi_\t}-1\right)\frac{\mathcal{F}_\t\mathcal{B}_\t}{u_\t^4}\cos^2\varphi\right)\right]^{-\frac{1}{2}}\,.
\end{split}
\end{equation}

\noindent
This is manifestly divergent. A convenient regularisation method is to 'subtract twice the mass of a single quark in the plasma'. The infinite mass $m_q$ of a single quark at zero temperature is simply given by 
\begin{equation}
 m_q = \frac{1}{2 \pi \alpha'} \int_0^{u_h} du\,\frac{\sqrt{\mathcal{B}}}{u^2}
\end{equation}

\noindent
The finite value for the binding energy of a $q\bar q$-pair in our anisotropic background is then given by
\begin{align}
  E &= - \left(\frac{S}{\mathcal{T}} + 2 m_q\right) \label{eq:freeEnergy} \\
    &= \frac{1}{\pi\alpha'} \left( \int_0^{u_\t} du \frac{\sqrt{\mathcal{B}}}{u^2\phantom{\Big\vert}} \left(\,\frac{1}{\sqrt{1 -\frac{\mathcal{F}_\t\mathcal{B}_\t u^4}{\mathcal{FB} u_\t^4}-\left(e^{\phi-\phi_\t}-1\right)\frac{\mathcal{F}_\t\mathcal{B}_\t u^4}{\mathcal{FB} u_\t^4}\cos^2\varphi}}\, - 1\right)  - \int_{u_\t}^{u_h}\frac{\sqrt{B}}{u^2\phantom{\Big\vert}} \right)\,.\nonumber
\end{align}

\noindent
This is the general formula for the binding energy for the anisotropic metric Ansatz \eqref{eq:OurMetricAnsatz}. The procedure now is to explicitly derive the metric functions in our model and use them in \eqref{eq:freeEnergy} to obtain the binding energy as a function of the parameters and the turning point $u_\t$. Then, together with the formulas \eqref{eq:getL1}--\eqref{eq:gettheta} for the distance and the angle $\theta$ of the $q\bar q$-pair we can derive the binding energy as a function $E(L,\theta)$. 

\begin{figure}[ht]\centering
 \includegraphics[width=.4\textwidth]{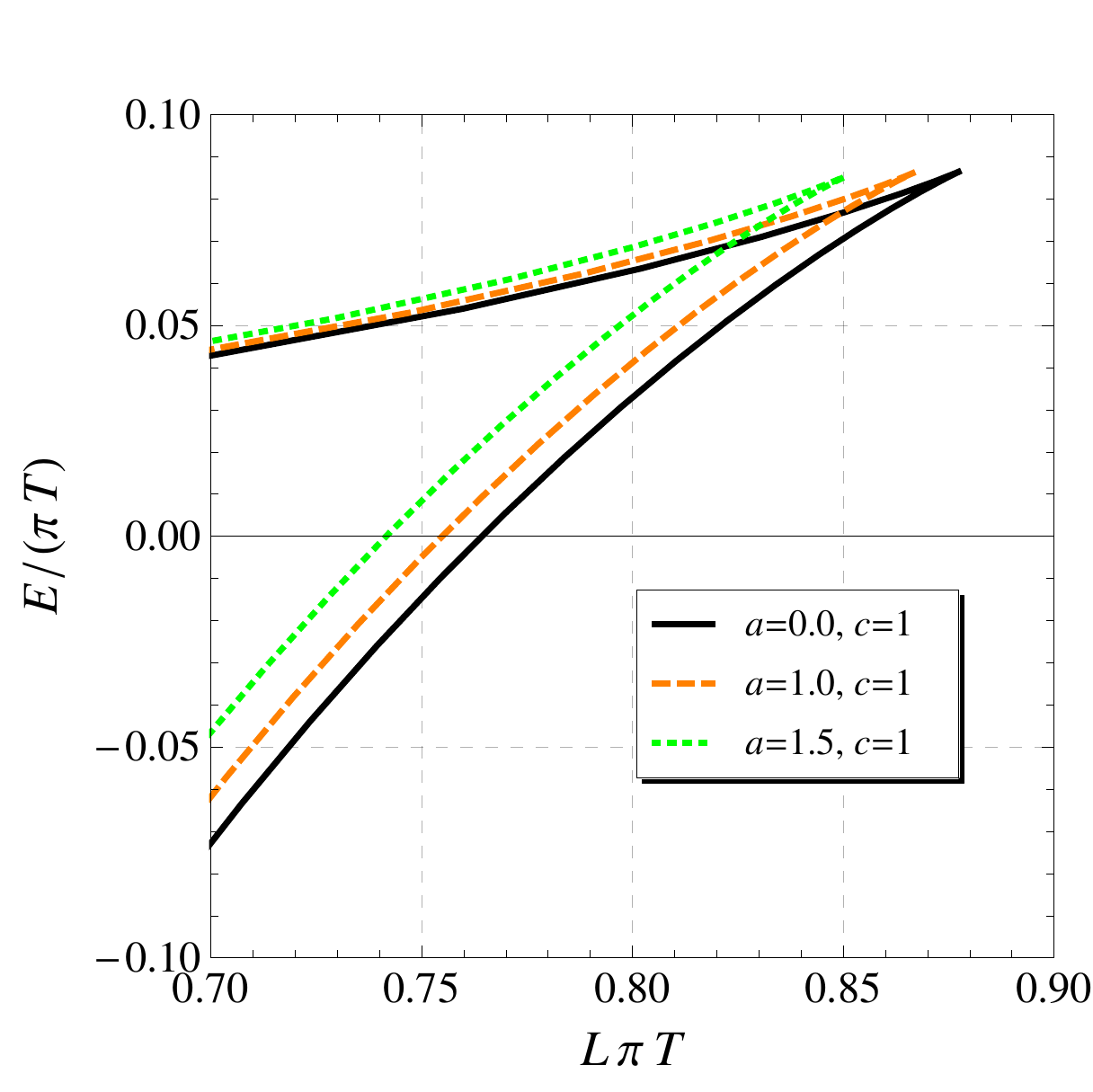}\hspace{30pt}
 \includegraphics[width=.4\textwidth]{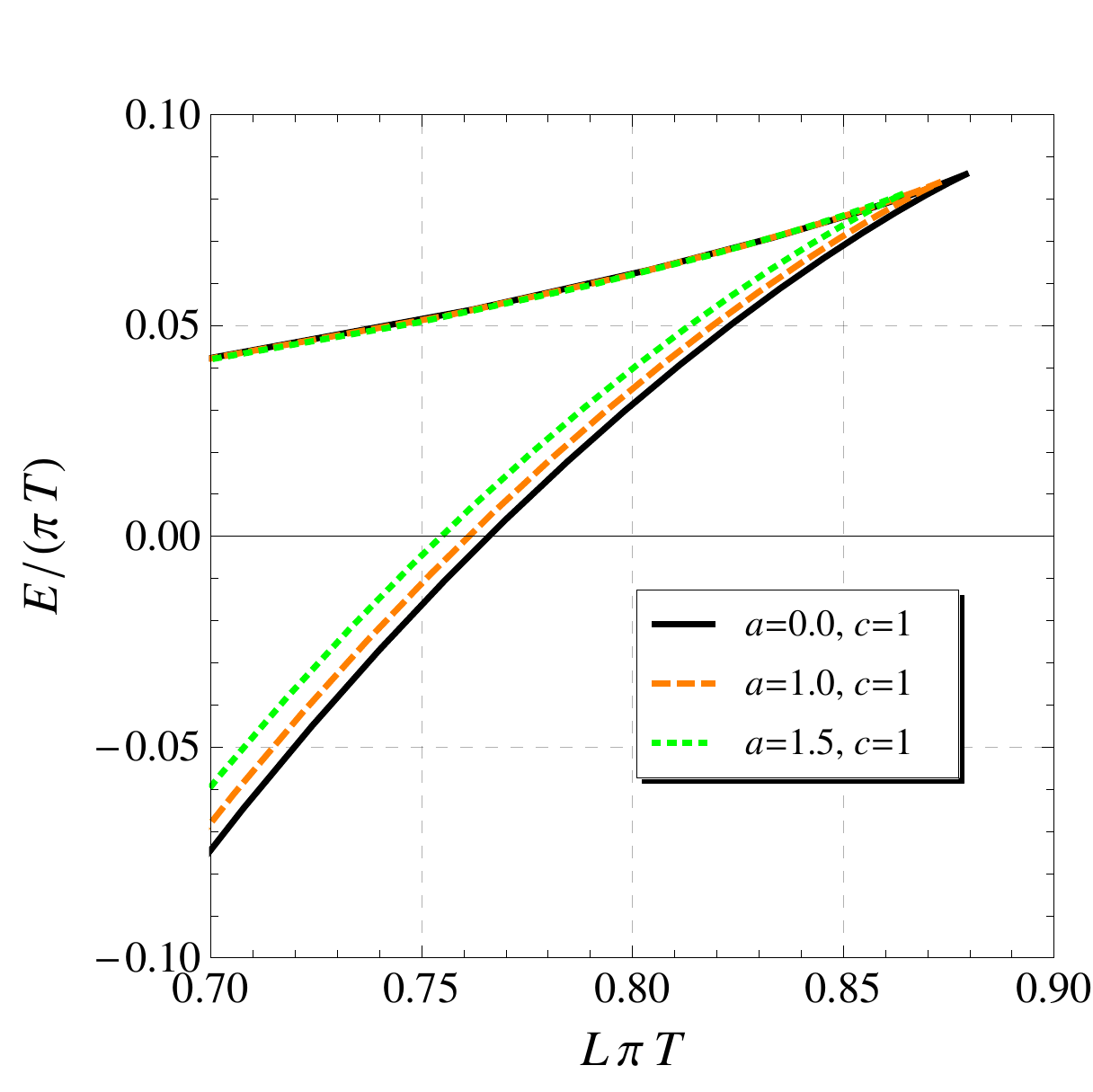}
 \caption{The binding energy of the $q\bar q$-pair aligned parallel (left) and perpendicular (right) to the anisotropic $z$-direction for different values of the deformation parameter $a$. We can again see that the screening distance shrinks which comes together with higher values of $E(L)$. We can also see that the effect of $a$ is much smaller for the perpendicular than for the parallel orientation, too.}
 \label{fig:EnerguiexaComb} 
\end{figure}

\begin{figure}[t]\centering
 \includegraphics[width=.4\textwidth]{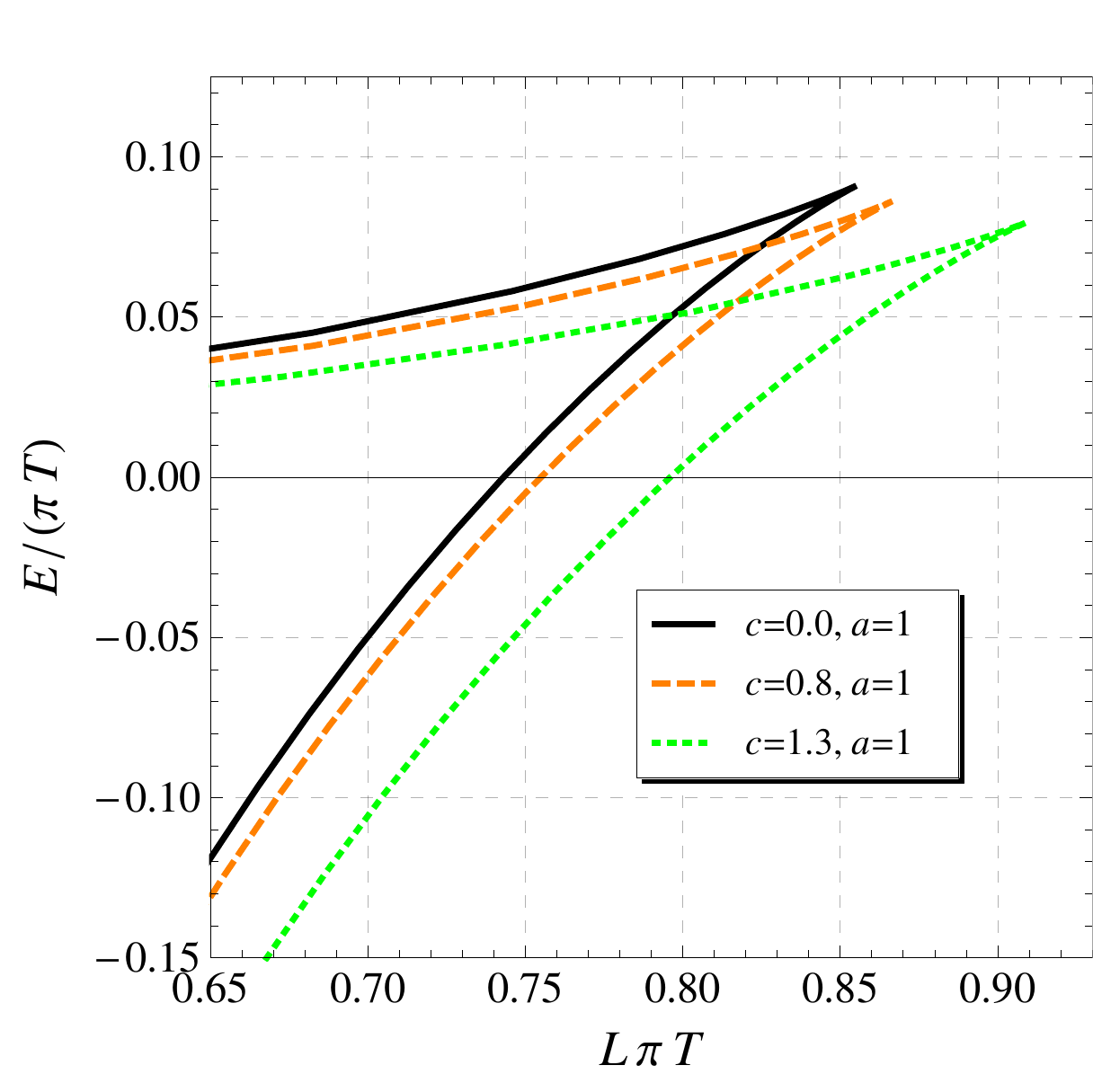}
 \hspace{30pt}
 \includegraphics[width=.4\textwidth]{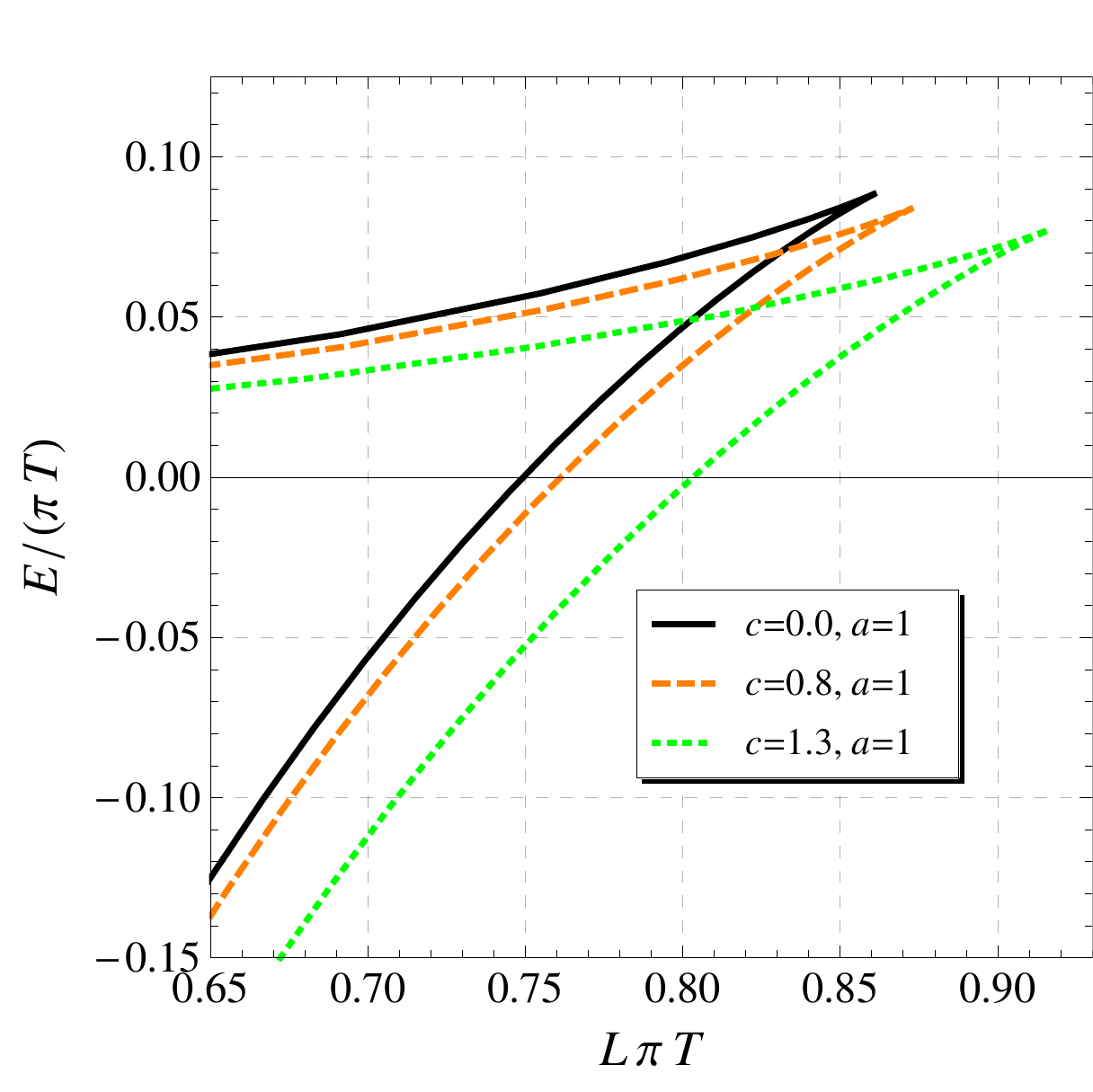}
 \caption{The binding energy of the $q\bar q$-pair aligned parallel (left) and perpendicular (right) to the anisotropic $z$-direction for different values of the deformation parameter $c$. We can again see that the screening distance grows with $c$ which comes together with lower values of $E(L)$. We can also see that the effect of $c$ is the same in the two orientations.}
 \label{fig:EnerguiexcComb} 
\end{figure}

Some numerical results for the two-parameter deformation are shown in figure \ref{fig:EnerguiexaComb} and \ref{fig:EnerguiexcComb}. At every possible distance of the $q\bar{q}$-pair there exist two possible string configurations but with different energies. It is understood that the configuration with the lower binding energy is the stable one. The first figure shows the influence of the anisotropic deformation on the binding energy for fixed isotropic deformation parameter $c$. The general behaviour does not change for different values of $c$, so that figure \ref{fig:EnerguiexaComb} is a representative for all $c$. We can see that for larger deformation parameter $a$ the binding energy grows. 

The second figure shows the influence of the isotropic deformation on the binding energy for fixed anisotropy parameter $a$. It is a representative for all anisotropic deformation parameters. For every fixed $a$ the qualitative influence of the deformation parameter $c$ is the same. We can see that the binding energy shrinks with $c$. The effect does almost not change parallel and perpendicular to the anisotropic direction. 

\subsection{Drag Force on a Single Moving Quark}

We now want to derive the drag force on a heavy quark moving with constant velocity through the anisotropic strongly coupled plasma. First we briefly want to show how to do this in an anisotropic geometry with a general metric Ansatz \eqref{eq:OurMetricAnsatz}.

Here, we closely follow the logic of \cite{Chernicoff:2012iq}, where one can also find a more detailed derivation. A simple model for the setting is described by the equation of motion 
\begin{equation}
 \frac{d \vec{p}}{dt} = -\mu \vec{p} + \vec{F}\,,
\end{equation}
where $\vec{p}$ is the quark's momentum, $\mu$ is the drag coefficient, and $\vec{F}$ is an external force. To keep a uniform motion, the necessary condition is $\vec{F} = \mu \vec{p}$. The dual description to a infinitely massive quark in the anisotropic plasma is an open string attached to the boundary at $u=0$ propagating in the anisotropic geometry. Its dynamics is governed by the Nambu-Goto action
\begin{equation} \label{eq:DragAction1}
 S = -\frac{1}{2 \pi \alpha'} \int_\Lambda d\iota\,d\sigma \,\sqrt{-g} \equiv \int_\Lambda d\iota\,d\sigma \,\mathcal{L}\,,
\end{equation}
where $\Lambda$ is the worldsheet of the string, $g$ is the determinant of the induced worldsheet metric, $\sigma$ is the spacelike and $\iota$ is the timelike coordinate on the worldsheet. Let us denote the spacetime coordinates by $X^M$, then the flow of spacetime momentum $\Pi_M$ along the string is given by 
\begin{equation}
 \Pi_M = \frac{\partial \mathcal{L}}{\partial(\partial _\sigma X^M)}\,.
\end{equation}

\noindent
The external force on the quark that is needed for the constant velocity may be gathered by attaching the endpoint of the string to a $D_7$-brane and turning on a constant electric field $F_{MN} = \partial_{[M}A_{N]}$ on the brane. Then we have to add to the Nambu--Goto action the boundary term
\begin{equation}
 S_\text{bdry} = -\int_{\partial \Lambda} d\iota\, A_N \partial_\iota X^N = - \frac{1}{2} \int_{\partial \Lambda} d\iota\,F_{MN}X^M\partial_\iota X^N. 
\end{equation}

\noindent
We demand that the boundary term arising from the variation of the total action $S + S_\text{bdry}$ vanishes. This yields the boundary condition 
\begin{equation}\label{eq:boundaryCondDrag}
 \Pi_M+ F_{MN} \partial_\tau X^N \vert_{\partial \Lambda} = 0\,.
\end{equation}

\noindent
In an anisotropic background we expect that the drag force depends on the direction of the velocity of the quark relative to the anisotropic direction. Thus, we introduce the angle $\theta$ between the velocity $\vec{v}$ of the quark and the $z$-axis. A suitable string embedding is then given by
\begin{align}
 x(t,u) = \Big(vt + \xi(u)\Big)\sin\theta\,,\\
 z(t,u) = \Big(vt + \zeta(u)\Big)\cos\theta\,,
\end{align}
which describes a quark moving with velocity $v$ in the $xz$-plane at angle $\theta$ with the anisotropic $z$-direction. With this the Nambu-Goto action takes the form 
\begin{align}
  S &\equiv \mathcal{T} \int du\,\mathcal{L}\,,
\end{align}
where 
\begin{align}
 \mathcal{L}&= -\frac{1}{2\pi\alpha'} \sqrt{ \frac{\mathcal{FB}+\sin^2\hspace{-2pt}\theta\left(\mathcal{F}^2\mathcal{B}\,\xi'^2-v^2\right) +e^{-\phi}\cos^2\hspace{-2pt}\theta\left(\mathcal{F}^2\mathcal{B}\zeta'^2-v^2-\mathcal{F}v^2\left(\xi'-\zeta'\right)^2\sin^2\hspace{-2pt}\theta\right)}{\mathcal{F}u^4}}\,.
\end{align}

\noindent
The rates of energy and momentum flow down the string towards the horizon are then given by
\begin{align}
 -\Pi_t&= \frac{\partial \mathcal{L}}{\partial x'}\partial_tx+ \frac{\partial \mathcal{L}}{\partial z'}\partial_tz = v\left(\Pi_x+\Pi_z\right) = \frac{1}{\mathcal{L}\,u^4}\mathcal{FB}v\left(\xi'\sin^2\theta + e^{-\phi} \zeta'\cos^2\theta\right),\\[1em]
 \Pi_x&= \frac{\partial \mathcal{L}}{\partial x'} = \frac{\sin \theta}{\mathcal{L}\,u^4} \left(\mathcal{FB}\,\xi'+e^{-\phi}v^2\left(\zeta'-\xi'\right)\cos^2\theta\right), \label{eq:Pixaniso}\\[1em]
 \Pi_z&=\frac{\partial \mathcal{L}}{\partial z'} = \frac{\cos \theta}{\mathcal{L}\,u^4}\, e^{-\phi}\Big(\mathcal{FB}\,\zeta'+v^2\left(\xi'-\zeta'\right)\sin^2\theta \Big), \label{eq:Pizaniso}
\end{align}
and the boundary conditions \eqref{eq:boundaryCondDrag} become
\begin{equation}
 \Pi_x = F_x\,,~~~\Pi_z = F_z\,,~~~-\Pi_t = F_x\,v\,\sin\theta+F_z\,v\,\cos\theta\,,
\end{equation}
with $(F_x,F_z)$ the two components of the external force. The first two equations show that the external force compensates the momentum loss by the quark into the medium and the third equation shows that the rate at which energy flows into the medium equals the work done by the external force. For positive energy and momentum flow from the boundary to the horizon, $0<\Pi_x,\Pi_z,-\Pi_t$, the solution to the above equations of motions is a string trailing behind the quarks position. We call this the \textit{physical} solution. 

We will now analyse the string profile and the corresponding energy and momentum flows for arbitrary $v,\theta$. We first notice that $\xi(u) \neq \zeta(u)$. Indeed, if $\xi(u) = \zeta(u)$ then the ratio of the two momenta would be given by
\begin{equation}
 \frac{\Pi_x}{\Pi_z} =  e^{\phi(u)} \, \tan \theta\,.
\end{equation}

\noindent
Since the left-hand side is constant and the right-hand side depends on $u$, this would be a contradiction. Thus, we indeed have $\xi(u) \neq \zeta(u)$ which means that the string does not trail `below' the trajectory of its endpoint. 

For the further analysis of the correct string shape we invert \eqref{eq:Pixaniso} and \eqref{eq:Pizaniso} as in \cite{Chernicoff:2012iq} to obtain
\begin{equation}
\label{eq:DragEOManiso}
 \begin{split}
  \xi' &= \pm \frac{e^{-\phi/2}\,v}{\mathcal{F}\sqrt{\mathcal{B}}}\,\frac{N_x}{\sqrt{N_xN_z -D\phantom{\vert}}}\, \\[1em]
  \zeta' &= \pm \frac{e^{\phi/2}\,v}{\mathcal{F}\sqrt{\mathcal{B}}}\,\frac{N_z}{\sqrt{N_xN_z -D\phantom{\vert}}}\,
 \end{split}
\end{equation}
where
\begin{align}
 N_x &= -\tilde{\Pi}_x \left(\mathcal{FB} \csc\theta -v^2 \sin\theta \right) +\tilde{\Pi}_z\,v^2\,\cos\theta \,,\\
 N_z &= -\tilde{\Pi}_z \left(\mathcal{FB} \sec\theta -e^{-\phi}v^2 \cos\theta \right) +\tilde{\Pi}_x\,e^{-\phi}\,v^2\,\sin\theta\,,\\
 D   &= \frac{\mathcal{FB}}{u^4\sin\theta\cos\theta}\left(\tilde{\Pi}_x\tilde{\Pi}_z\,u^4-e^{-\phi}\,v^2\sin\theta\cos\theta\right)\left(\mathcal{FB}-v^2\left[e^{-\phi}\cos^2\theta+\sin^2\theta\right]\right),\label{eq:D}
\end{align}
with $\tilde\Pi_{x,z} = 2\pi\alpha' \Pi_{x,z}$. We require that the solutions are real for all $0<u<u_\text{h}$.  Thus, the radicand $N_xN_z-D$ in \eqref{eq:DragEOManiso} must not be negative. This requirement leads us to definite values for $\tilde{\Pi}_x$ and $\tilde{\Pi}_z$. In our model $\mathcal{FB}$ is monotonically decreasing from 1 at
the boundary to 0 at the horizon, whereas $e^{-\phi}$ is increasing from the boundary to the horizon. Thus, the last factor in \eqref{eq:D} is positive at the boundary and negative at the horizon and there exists a value $u_\c$ in between such that
\begin{equation}
 \mathcal{F_\c B_\c} - v^2\left[e^{-\phi_\c}\cos^2\theta+\sin^2\theta\right] = 0\,, \label{eq:DragUc}
\end{equation}
where a subscript c stands for the function evaluated at $u_\c$. For this value $D$ vanishes and we get
\begin{equation}
 N_x N_z\vert_{u_\c} = -v^4 \left(e^{-\phi_\c}\, \tilde{\Pi}_x\cos\theta-\tilde{\Pi}_z\sin\theta\right)^2\,,
\end{equation}
which is negative unless the quotient of the rates of the momentum fluxes are given by
\begin{equation}
 \frac{\tilde{\Pi}_x}{\tilde{\Pi}_z} = e^{\phi_\c}\tan\theta. \label{eq:DragFirstCondition}
\end{equation}

\noindent
At last, also the first factor in brackets in \eqref{eq:D} exhibits a zero, since it is negative at the boundary but positive at the horizon. For the real solution it should vanish for the same $u = u_\c$. The latter condition, together with \eqref{eq:DragFirstCondition}, restricts the values of the two rates to be
\begin{equation}
 \tilde{\Pi}_x = \frac{v \sin\theta}{u_\c^2}\,,~~~~\tilde{\Pi}_z = e^{-\phi_\c}\frac{v\cos\theta}{u_\c^2}\,, \label{eq:PiSol}
\end{equation}
for a real solution from the boundary to the horizon. 

With \eqref{eq:PiSol} in hand we have obtained the force $\vec{F} = (\Pi_x,\Pi_z)$ that has to act on the quark in order for it to stay in uniform motion. In terms of the quark's velocity $\vec{v} = v(\sin\theta,\cos\theta)$ the force is given by
\begin{equation}
 \vec{F} = \frac{1}{2 \pi \alpha'}\,\frac{v}{u_\c^2} \left(\sin\theta\,,\,e^{-\phi_\c}\cos\theta\right)\,. \label{eq:Faniso}
\end{equation}

\noindent
It is worth mentioning that the velocity and the force that acts on the quark to stabilise the velocity are not aligned with one another except in the isotropic case, where $\phi_\c = 0$, or when the velocity is precisely along the anisotropic or isotropic direction, where $\theta = 0,\pi/2$, respectively. In the following we call the angle between the velocity and the force $\Delta\theta_F$. Note that the direction of the force changes for different $v$ (but constant $\theta$), too. This is because it appears in \eqref{eq:DragUc} and, thus, also determines $u_\c$ and in the end $\phi_\c$. 


\begin{figure}[t]\centering
 \includegraphics[width =.7\textwidth]{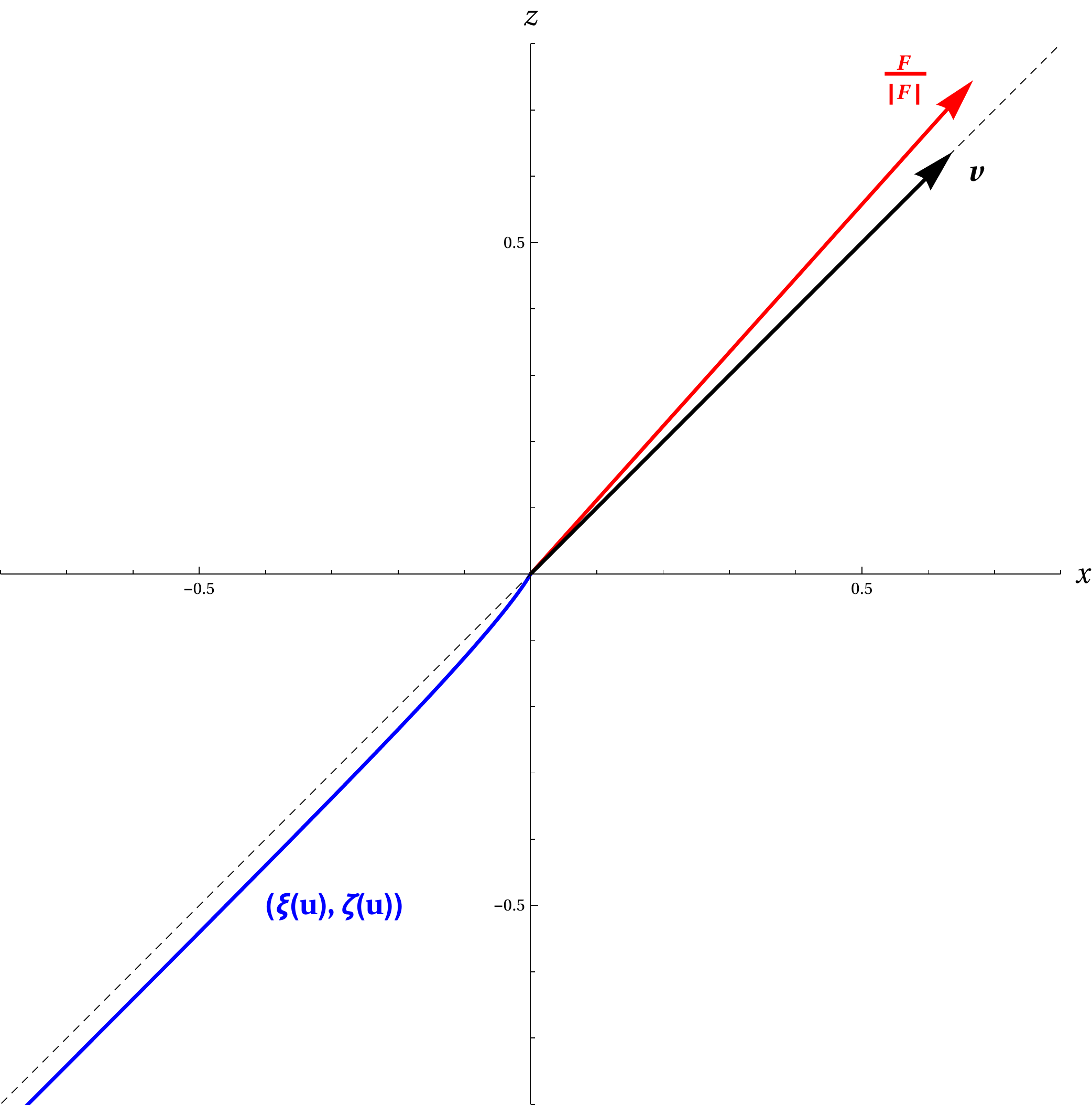}
 \caption{Projection of the string configuration onto the $xz$-plane and the corresponding force on the quark in the anisotropic background for a velocity $v = 0.9$, an angle of $\theta = 45^\circ$, $a/(\pi T)=2$, and $c=1$.}
 \label{fig:StringConfAniso} 
\end{figure}

Now, we have all the ingredients we need to calculate the drag force and the string profile for a given velocity $\vec{v} = v(\sin\theta,\cos\theta)$. The general approach for a given background with the metric functions $\mathcal{F,B}$ and the dilaton $\phi$ is:
\begin{enumerate}
 \item Solve \eqref{eq:DragUc} to obtain the critical point $u_\c$ for given $v$ and $\theta$. 
 \item Use $u_\c$ in \eqref{eq:PiSol} to obtain the rates of the momenta $\Pi_x,\Pi_z$ and the force $\vec{F}$ from \eqref{eq:Faniso} that is needed for a uniform motion. The drag force is then $\vec{F}_\text{drag} = -\vec{F}$.
 \item Use $\Pi_x,\Pi_z$ for the integration of \eqref{eq:DragEOManiso} to obtain the string profile.  
\end{enumerate}

\noindent
We want to present some numerical results for the drag force. We begin with figure \ref{fig:StringConfAniso} where we show the projection of the string configuration on the $xz$-plane together with the corresponding force to maintain a constant velocity $v=0.9$ at an angle of $\theta=45^\circ$, for a deformation parameter $a/(\pi T)=2$. 

For any $a\neq 0$ both the force and the tangent vector at $u=0$ ($\Leftrightarrow \xi=0=\zeta$) will not align with the velocity. The force always points more towards the anisotropic direction than the velocity.  The same holds for the tangent vector of the projection of the string profile. Thus, the string trails towards the anisotropic direction.  However, for $u\rightarrow u_\text{h}$ ($\Leftrightarrow x,z\rightarrow \infty$) the tangent vector and the velocity are becoming aligned. It is not too surprising that the force and the velocity are not aligned for an anisotropic background. In the gauge theory we expect that the pressure in the two directions are unequal. A naive expectation would then be that the forces to hold the velocities in these directions depends on that pressure. For the simple case $v_x = v_y$ it then follows that $F_x\neq F_y$. This is exactly what we can see in our result. 

\begin{figure}[ht]\centering
  \includegraphics[width = .49\textwidth]{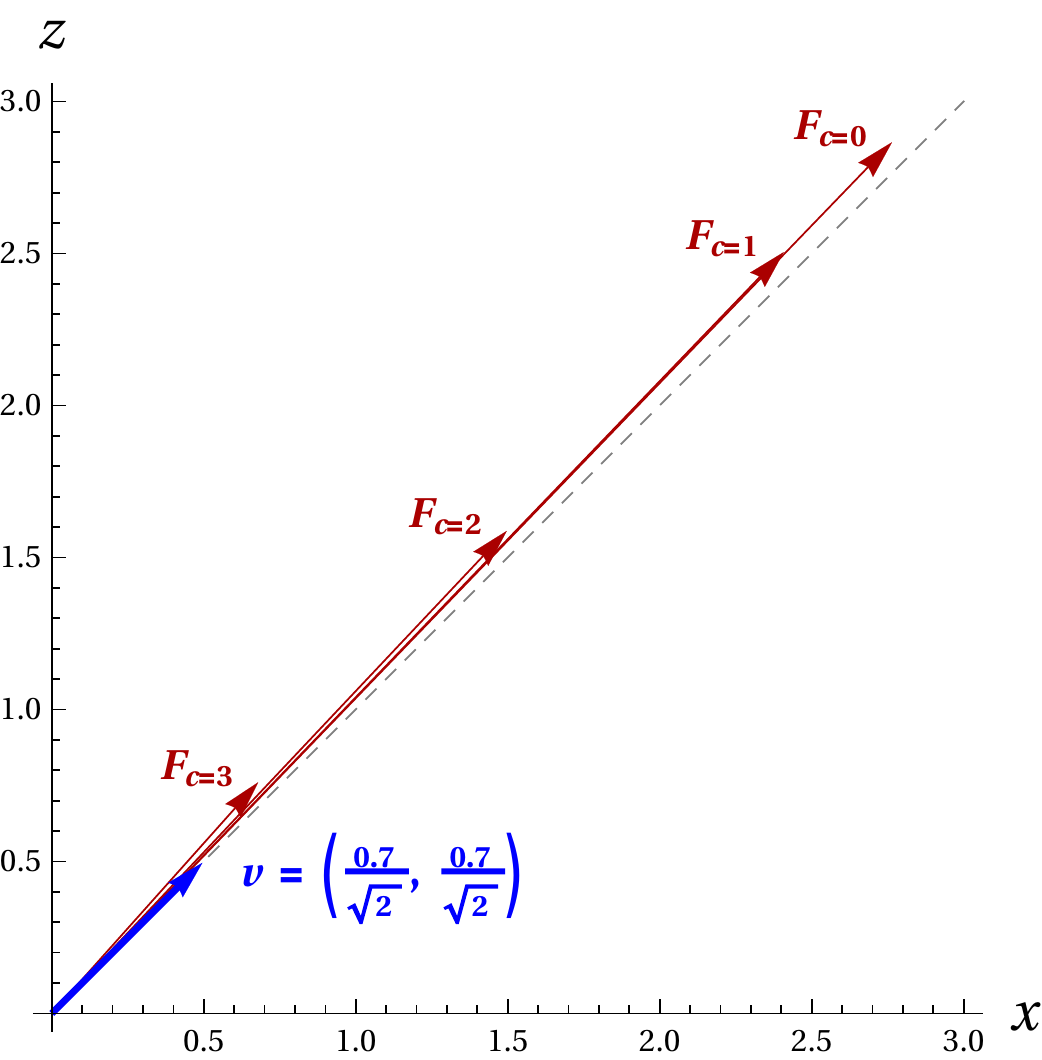}
  \includegraphics[width = .49\textwidth]{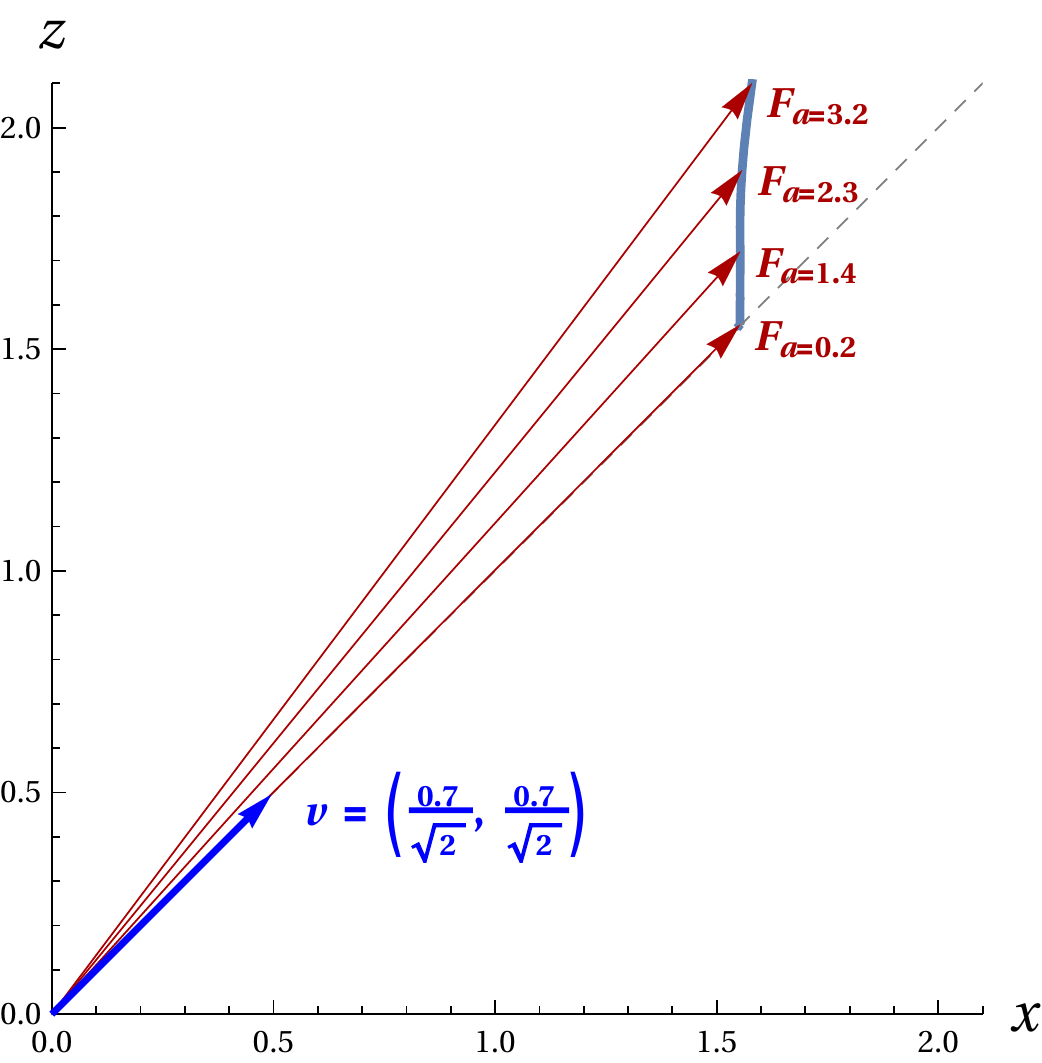}
  \caption{It is shown how the force changes for different values of the two deformation parameters $c$ and $a$ for a velocity $v=0.7$ at an angle $\theta = 45^\circ$. For the left plot, the anisotropy parameter was taken to be $a=1.4$. The absolute value is decreasing whereas $\Delta\theta_F$ is increasing for growing $c$. For the right plot, the non-conformal deformation was chosen to be $c=1$.  However, the results look alike for any $c$. Both the absolute value of the force and $\Delta\theta_F$ increase for growing $a$.}
  \label{fig:DragxcAniso} 
\end{figure}

In the following we show how the velocity and the two parameters affect both the absolute value and also the direction of the force. In figure \ref{fig:DragxcAniso} we show how the force depends on the two deformation parameter $c$ and $a$ for a velocity $v=0.7$ at an angle $\theta = 45^\circ$ with the $z$-direction where the respective other parameter is chosen to be fixed. For growing $c$ the absolute value of the force is decreasing whereas the angle between the velocity and the force is increasing. For larger values of $a$ both the absolute value of the force and its angle to the anisotropic direction increase. The result for the angle is expected, if we turn back to the naive explanation  of the last paragraph. Larger $a$ means a larger anisotropy in the five-dimensional geometry which means we expect al larger anisotropy in the gauge theory.

\begin{figure}[ht]\centering
 \begin{tikzpicture}
 \node at (0,0) {\includegraphics[width = .57\textwidth]{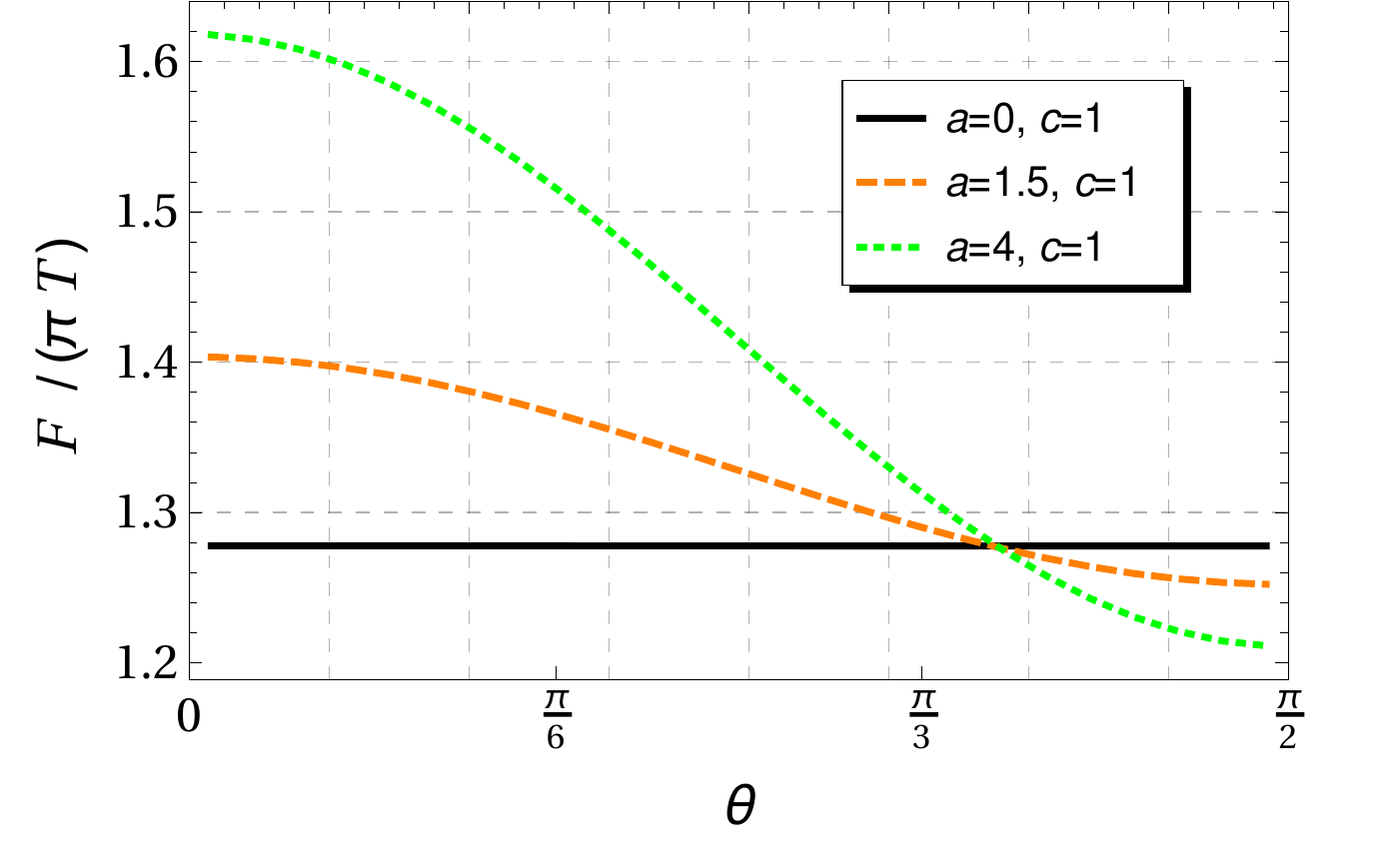}};
 \node at (2,-.5) {$\theta_\text{c}$};
\end{tikzpicture}
 \caption{We see that the effect of $a$ on the drag force depends on $\theta$. $\theta_c$ is a critical value where the drag force does not change at all.}
 \label{fig:DragxaAniso} 
\end{figure}

\begin{figure}[ht]\centering 
 \includegraphics[width = .49\textwidth]{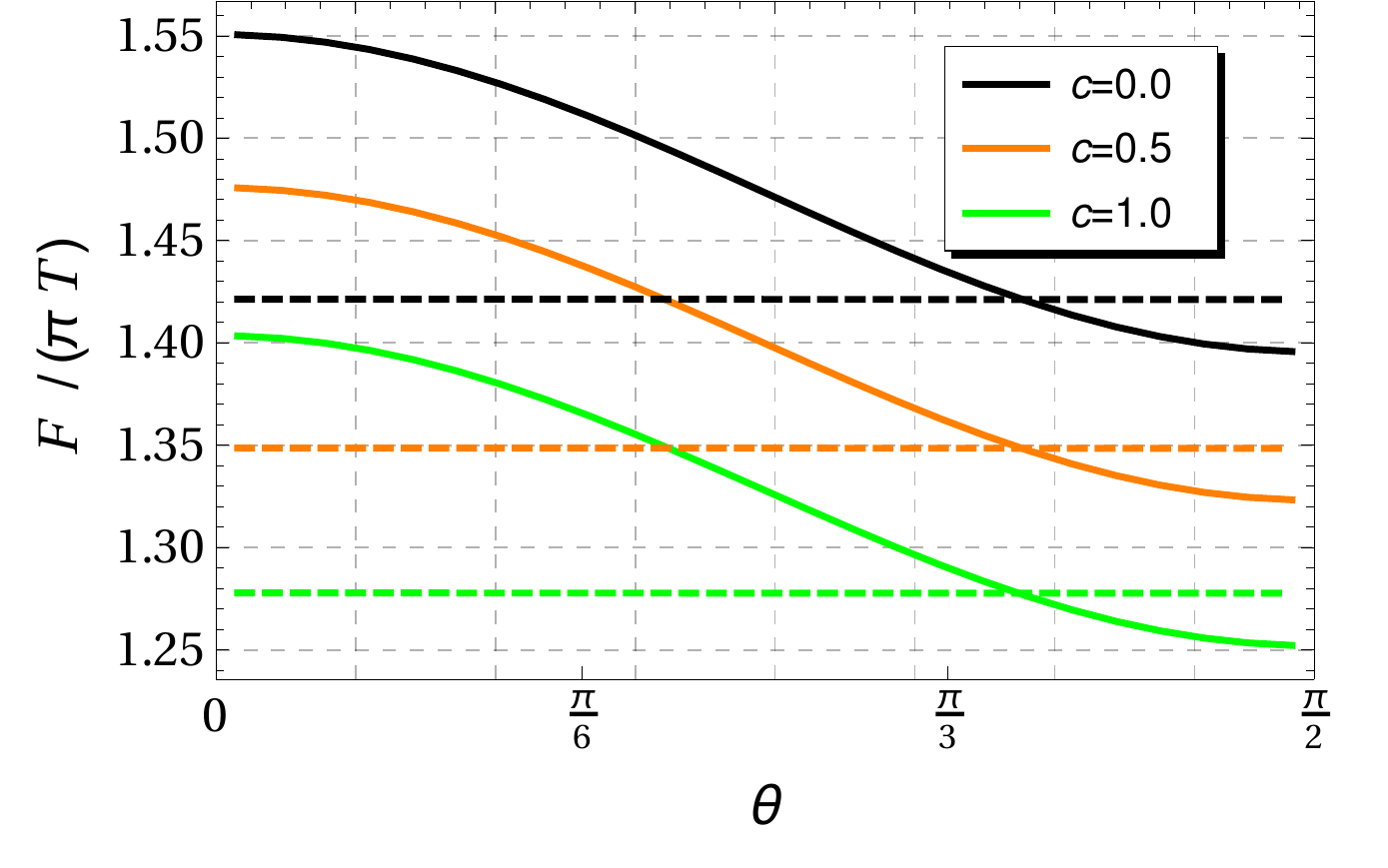}
 \includegraphics[width = .49\textwidth]{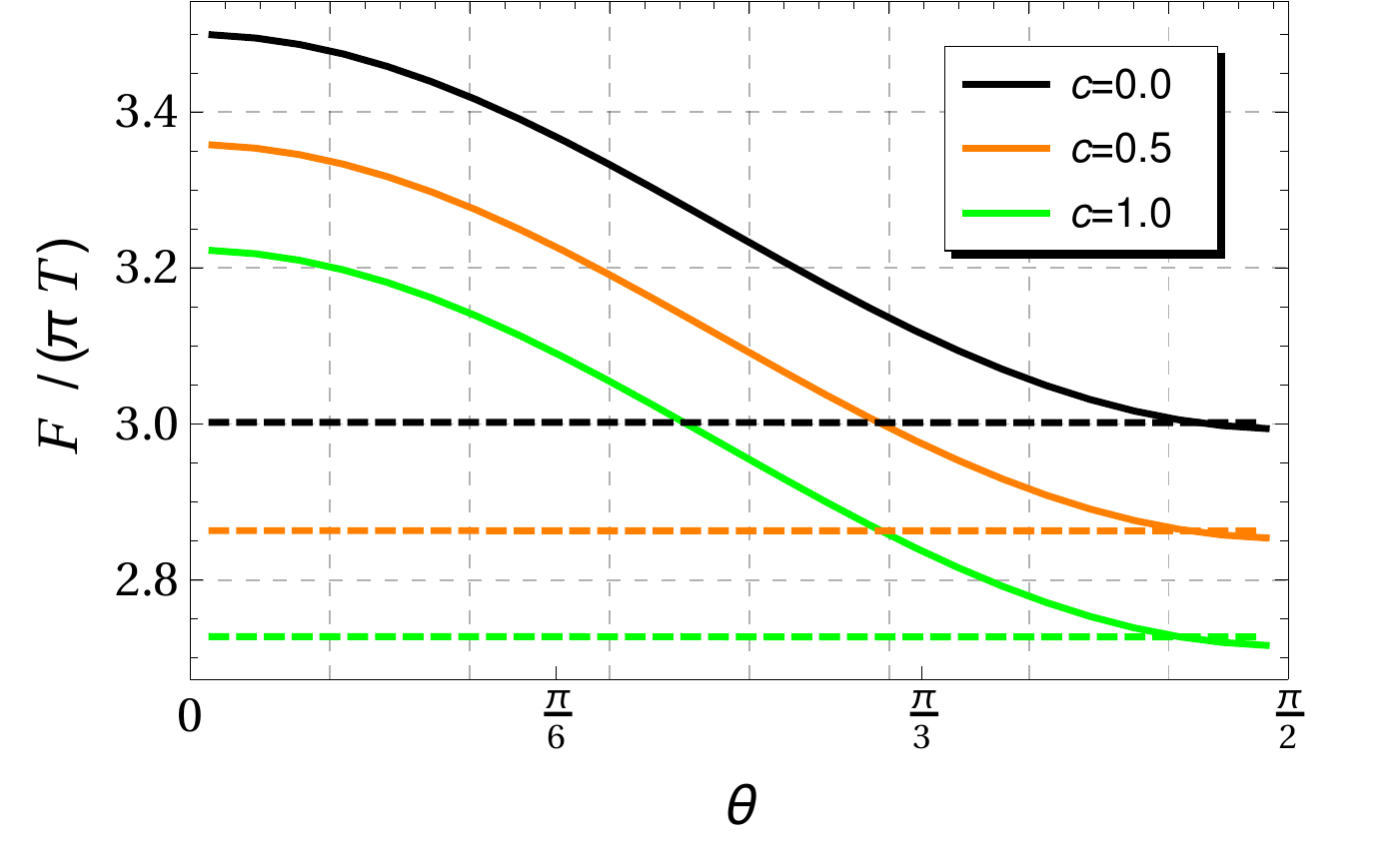}
 \caption{The drag force for different values of the non-conformal deformation parameter $c$ at $v=0.7$ (left) and $v=0.9$ (right). The horizontal lines are the drag force for $a/(\pi T)= 0$ and the curved lines are the drag force for $a/(\pi T) = 1.5$. The critical value $\theta_\c$ does not depend on $c$ but on the absolute value of the velocity $v$.}
 \label{fig:Fxtheta2} 
\end{figure}

The growth for the absolute value of the drag force for increasing $a$ is not generally true. This is visible in figure \ref{fig:DragxaAniso}. There we can see that the drag force only grows for angles $\theta$ smaller than a critical angle $\theta_\c$. It also shows that the drag force for $a\neq0$ has a directional dependants, as expected. Figure \ref{fig:Fxtheta2} shows exemplary that the critical angle $\theta_\c$ depends on the velocity but not on the 'non-conformal' deformation parameter $c$. 

Once again we can see the difference between the isotropic deformation (changing $c$) and the anisotropic deformation (changing $a$). The force changes completely differently for anisotropic and isotropic deformation. In a more mathematical language the behaviour of the absolute value of the drag force is given by 
\begin{align}
 \partial_c \vert \vec{F} \vert < &0  ~~~\forall_\theta\,,\\[1em]
 \partial_a \vert \vec{F} \vert >0 ~~~\forall_{\theta<\theta_\text{c}(v)} ~~~~&\text{and}~~~~\partial_a \vert \vec{F} \vert < 0 ~~~\forall_{\theta>\theta_\c(v)} \,.
\end{align}

\noindent
A very interesting feature is that for $\theta = \theta_\c$ the absolute value of the drag force does not change for different values of $a$ and is hence independent of the anisotropy of the plasma. However, the angle $\Delta \theta_F$ will still change. 

\section{Summary, Interpretation and Outlook} \label{sec:summary}

Let us first recall the motivation that has driven us. Our main interest lies in the description of the quark-gluon plasma that can be produced in heavy ion collisions. We think that the gauge/string duality can help to gain more insight into how matter behaves under these extreme conditions. 
The quark-gluon plasma produced in heavy ion collisions is manifestly anisotropic. It seems likely that the anisotropy itself affects observables and the general behavior of the plasma. When we want to find a proper description of the QGP through the gauge/string duality we have to consider anisotropy on both sides of the duality. In this work we considered models that are not fully anisotropic but posses one anisotropic direction. 

Our first achievement is the consistent construction of a two-parameter deformation of \SYM~that allows us to consider a larger class of anisotropic and non-conformal models that are specified by two parameters that we called $(c,a)$. 

We now want to summarize the results for the observables that we have studied. 

The screening distance $L_\sc$ of a static $q\bar q$-pair in the plasmas which is defined as the maximal possible distance at which the quark and the antiquark form a bound state is minimal for vanishing non-conformal deformation and grows monotonically for larger deformations, i.e. without anisotropy it is minimal in \SYM. In \cite{Ewerz:2013sma} this was shown to be true for general small isotropic deformations of \SYM. One may conjecture that the screening distance in \SYM~is a lower bound for all isotropic theories.\footnote{However, by introducing a chemical potential $\mu$ in an isotropic deformation one can get below the value of $L_\sc$ in \SYM~\cite{Samberg:2012}.} 

In the anisotropic theories, the screening distance shows the opposite behaviour and, in addition, depends on the orientation of the $q\bar q$-pair relative to the anisotropic direction. The effect of $a$ is strongest if the pair lies parallel to the anisotropic direction. However, even for a perpendicular orientation the anisotropy affects the screening distance. In a naive gauge theory picture this is easy to understand. A bound $q\bar q$ state is that of a cloud of virtual gluons between the two massive particles. The cloud extends in every direction even if the pair lies perpendicular to the anisotropic direction. In this picture the anisotropy always affects the gluon cloud and, thus, the anisotropy affects the screening distance for every orientation. 

The second observable we analyse for a static $q\bar q$-pair is its binding energy $E$ which reflects the strength of the bound state in the hot plasma. We compute the binding energy as a function of the distance of the quark to the antiquark.

For the isotropic non-conformal deformation the binding energy decreases for larger deformation and has a maximum for $c=0$. Thus, the bound state in the thermal background of the plasma gets strengthened for larger deformations. This is in agreement with the increase of the screeneing distance $L_\sc$. Under the anisotropic deformations the binding energy shows the opposite behavior. It increases for larger anisotropic deformations and, thus, the bound state in the hot plasma is weakened. It also depends on the orientation w.r.t. the anisotropic direction of the hot plasma. The effect is strongest parallel to the anisotropic direction. 

In the parameter regime we analyze the deviation from the values in \SYM~is smaller than $30\%$ for both observables. This may let us hope that the values of the screening distance and the binding energy are not too different from the values in the real anisotropic QGP.  

Our last observable is the drag force on a single quark moving with constant velocity. For non-conformal deformations the drag force decreases under deformations. In a very naive picture of the plasma this again matches the results of the screening distance and the binding energy of the static $q\bar q$-pair. All three behaviors can be explained if we assume that the plasma gets dilute for non-conformal deformations. 

The results again differ for the anisotropic deformations. If the velocity does not point either parallel or perpendicular to the anisotropic direction, neither the drag force nor the projection of the string configuration align with the velocity of the quark. This is expected for an anisotropic medium. Surprisingly, the dependency of the absolute value of the drag force on the anisotropy parameter $a$ depends on the direction and the absolute value of the static velocity of the quark. It can both shrink or grow and hence does not show a general pattern under anisotropic deformations. 

In a nutshell one can say that all analyzed observables behave differently under isotropic but non-conformal and anisotropic deformations. The behaviors of the observables also show that in our two-parameter deformation the two parameters $(c,a)$ really represent the deformations they are designed to represent. Thus, in our model $c$ controls isotropic non-conformal deformation and $a$ is the parameter controlling the anisotropy. 

In the anisotropic one-parameter deformation, \cite{Mateos:2011tv} interpret the resulting theory as an anisotropic version of \SYM. We want to adopt this interpretation for our two-parameter deformation. For fixed parameter $c$ the parameter $a$ parametrizes the anisotropy within the \textit{same} lower-dimensional gauge theory. The anisotropy is then for example represented by an anisotropic stress-energy tensor of that gauge theory. 

If this interpretation is indeed correct, the different results for isotropic and anisotropic deformations are very natural. The parameter $c$ which controls the non-conformal deformation changes the dual gauge theory whereas for different $a$ the underlying gauge theory stays the same but corresponds to an anisotropy of the plasma. They really are two different types of deformation and, thus, different results for the behavior of observables under change of these is natural.

With this work we think we went a small step towards a better understanding of the QGP as it is produced in heavy ion collisions. We enlarged the class of anisotropic models and, thus, can investigate features of a larger class of anisotropic strongly coupled plasmas. After one step there follows the next one. In this case this could mean to enlarge the class of models further. One could for example imagine an anisotropy in all the spatial directions. So far, there does not exist an Ansatz in gauge/string duality which handles full anisotropy. Another idea is to introduce a chemical potential similar to the approaches in \cite{DeWolfe:2010he,Samberg:2012,Cheng:2014sxa} as attempted in \cite{Cheng:2014qia} or one can explicitly investigate breaking of supersymmetry as {e.g.} in \cite{Chakraborty:2017wdh}. 

In this work we analysed several observables but it is clear that there are many more that can be investigated. One can have a look at the running coupling or rotating quarks as for example in \cite{Ewerz:2013sma,Fadafan:2008bq}. One could also have a closer look at the stress-energy tensor and possible instabilities as in the anisotropic one-parameter deformation in \cite{Mateos:2011tv}. In principle any observable that has been investigated in anisotropic deformations \cite{Wang:2016noh,Ali-Akbari:2014xea,Giataganas:2013lga,Chakraborty:2012dt,Mandal:2012wi,Chakrabortty:2013kra,Jahnke:2013rca,Ge:2014aza,Fadafan:2013bva,BitaghsirFadafan:2017tci} can now also be considered in the two parameter model. It might also be interesting to check whether the present model shows confing as the anisotropic models in \cite{Giataganas:2017koz,}. It might also be interesting to consider (quantum) information theoretical aspects \cite{Brown:2015bva,Pastawski:2015qua,Brehm:2016dzg} and in particular entanglement \cite{Ryu:2006bv} from the holographic perspective and see how non-conformal and in particular anisotropic deformations affect them. 

Our final conclusion is that the gauge/string duality is a valuable tool for understanding the strongly interacting anisotropic QGP as it is produced in heavy ion collisions. Although \SYM~and (anisotropic) deformations thereof are surely not QCD, it may provide some insight into the structure of strongly coupled quantum field theories and their (anisotropic) plasmas. We are certain that the gauge/string duality can lead to a better understanding of QCD and its different phases. 

\newpage
\appendix
\section{\boldmath Derivations of $T$}
\label{app:TandS}

To calculate the temperature first replace $t\rightarrow -it_\text{E}$ in \eqref{eq:IsoMetric}, where $t_\text{E}$ is then the Euclidean time coordinate and perform a leading order expansion around the black brane position $u_\text{h}$:
\begin{equation}
 ds_\text{E}^2 \approx \frac{1}{u_\text{h}^2} \left(\mathcal{F}'_\text{h} \mathcal{B}_\text{h} (u_\text{h} -u)dt_\text{E}^2 +d\vec{x}^2 + \frac{du^2}{\mathcal{F}'_\text{h} (u_\text{h}-u)}\right),
\end{equation}
where a prime denotes a derivative with respect to $u$ and a subscript h means that the function is evaluated at $u=u_\text{h}$. We assume that $\mathcal{B}'_\text{h} = 0$. The latter can be rewritten as
\begin{equation}
 ds_\text{E}^2 \approx \frac{1}{u_\text{h}^2} \left( \rho^2d\theta^2 +d\rho^2+d\vec{x}^2  \right),\label{haha}
\end{equation}
with
\begin{equation}
  \label{haha2}
 \rho = 2 \sqrt{\frac{(u-u_\text{h})}{\vert \mathcal{F}'_\text{h}\vert}} ~~~~~ \text{, and}~~~~~~ \theta = \frac{1}{2}\vert \mathcal{F}'_\text{h}\sqrt{\mathcal{B}_\text{h}} \vert t_\text{E}. 
\end{equation}

\noindent
The first two terms in the metric \eqref{haha} describe a plane in polar coordinates, so in order to avoid a canonical singularity at $\rho = 0$ we must require $\theta$ to have a period $2\pi$. From \eqref{haha2} we then see that the period of the Euclidean time which is interpreted as $\gamma = 1/T$ must be 
\begin{equation}
 \gamma = \frac{1}{T} = \frac{4\pi}{\vert \mathcal{F}'_\text{h}\sqrt{\mathcal{B}_\text{h}} \vert}
\end{equation}

\noindent
The latter argumentation is true for the anisotropic metric Ansatz \eqref{eq:OurMetricAnsatz}, too. 

\section{Long equations}
\label{app:equ}
The third order differential equation for the scalar $\tau$ of section \ref{sec:deform1} is given by
\begin{equation}\label{eq:tauDGL2}
 \begin{split}
   0=& ~c\, u\, \tau ^4 \left(18 \tau'^2+u^2 \tau'^4+u^3 \tau'^3 \tau ''-12 u^2 \tau''^2+6 u
\tau ' \left(3 \tau ''+u \tau ^{(3)}\right)\right)+\\
&+12 u \left(24 \tau'^2+u^2 \tau'^4+u^3 \tau'^3 \tau ''-12
u^2 \tau''^2+6 u \tau ' \left(4 \tau ''+u \tau ^{(3)}\right)\right)-\\
&-36 \tau  \left(10 \tau '+u^2 \tau'^3+u^3 \left(\tau
'\right)^2 \tau ''+2 u \left(7 \tau ''+u \tau ^{(3)}\right)\right)-\\
&-36 c \tau ^3 \left(10 \tau '+u^2 \tau'^3+u^3 \tau'^2
\tau ''+2 u \left(7 \tau ''+u \tau ^{(3)}\right)\right)+\\
&+2 u \tau ^2 \bigg(18 (1+6 c) \tau'^2+u^2 \tau'^4+u^3 \tau'^3 \tau ''-12 u^2 \tau''^2+\\
&~~~~~~~~~~+6 u \tau ' \left(3 (1+6 c) \tau ''+u \tau ^{(3)}\right)\bigg)\,.
 \end{split}
\end{equation}
where a prime denotes a differentiation with respect to $u$. 

The third order diff. equ. for the dilaton $\phi$ in the single parameter anisotropic deformation of section \ref{sec:MT} is given by
\begin{small}
 \begin{equation}
 \begin{split}
0  = & ~a^2 u e^{7 \phi /2} \bigg\{13 u^3 \left(\phi '\right)^4+8 u
   \left(\phi '' \left(11 u^2 \phi ''-60\right)-12 u \phi
   ^{(3)}\right)+u^2 \left(\phi '\right)^3 \left(13 u^2 \phi
   ''+96\right)+\\
 &+2 u \left(\phi '\right)^2 \left(-5 u^3 \phi
   ^{(3)}+53 u^2 \phi ''+36\right)+\phi ' \bigg(30 u^4 \left(\phi
   ''\right)^2-32 \left(2 u^3 \phi ^{(3)}+9\right)+32 u^2 \phi
   ''\bigg)\bigg\}+\\
 &+32 \Big\{-u^2 \phi ^{(3)} \phi ' \left(5 u \phi
   '+12\right)+3 u^2 \left(\phi ''\right)^2 \left(5 u \phi
   '+8\right)-\left(\phi '\right)^2 \Big(u \phi ' \left(11 u \phi
   '+30\right)+36\Big)-\\
 & -u \phi ' \phi '' \Big(u \phi ' \left(11 u
   \phi '+25\right)+36\Big)\Big\}\,.  
 \end{split}
 \label{eq:phiDGLMT}
\end{equation}
\end{small}

\noindent
The differential equations for the scalar $\tau$ and the dilaton $\phi$ in the two-parameter model are given by
\begin{align}
\label{eq:combDGL1} 0 = &\, 4 \phi ' \left(12 \left(\tau +c \tau ^3\right)+u^2 \left(12+2 \tau ^2+c \tau ^4\right) \tau ''\right)-4 u \left(-12 \left(\tau
 +c \tau ^3\right)+\right.\\
\nonumber &+ \left. u \left(12+2 \tau ^2+c \tau ^4\right) \tau '\right) \phi ''-3 a^2 e^{7 \phi /2} u^2 \left(-u \left(4+u \phi '\right) \tau ''+\tau
' \left(-4+u^2 \phi ''\right)\right).
\end{align} 
\begin{small}
\begin{align}\label{eq:combDGL2}
 0 = &  \left(4 \phi ' \left(8 \left(\tau +c \tau ^3\right) \tau '-\left(12+2 \tau ^2+c
\tau ^4\right) \phi '\right)+8 \tau ^2 \left(2+c \tau ^2\right) \phi '' +24 \left(a^2 e^{7 \phi /2}+4 \phi ''\right)\right. \\
\nonumber &+\left. 6 a^2 e^{7 \phi /2} u \left(\phi
' \left(14+3 u \phi '\right)+u \phi ''\right)\right)u \left(12+5 u \phi '\right) \left(\phi '+u \phi ''\right)-\left(\phi '+u \phi ''\right)  \times \\
\nonumber&\times \left(4 \left(12+2 \tau ^2+c \tau ^4\right) \left(\phi ' \left(72+4
u^2 \left(\tau '\right)^2+u \phi ' \left(48+17 u \phi '\right)\right)-4 u \left(6+5 u \phi '\right) \phi ''\right)+\right.\\
\nonumber&+ \left. 3 a^2 e^{7 \phi /2} u \left(384+u
\left(4 u \left(\tau '\right)^2 \left(4+u \phi '\right)+\phi ' \left(304+u \phi ' \left(116+17 u \phi '\right)\right)-\right.\right.\right.\\
\nonumber &\left.\left.\left.-4 u \left(16+5 u \phi '\right)\phi ''\right)\right)\right)-2 \left(4 \left(12+2 \tau ^2+c \tau ^4\right) \phi '+3 a^2 e^{7 \phi /2} u \left(4+u \phi '\right)\right) \left(2 \phi
''+u \phi ^{(3)}\right),
\end{align}
\end{small}

\acknowledgments

I like to thank Carlo Ewerz for initializing and supervising this project. I also thank Andreas Samberg for useful discussions on the subject.

\bibliographystyle{JHEP}
\bibliography{Literatur}

\end{document}